\documentclass[onecolumn]{aastex63}
\usepackage{textcomp}
\usepackage{graphicx}
\usepackage{amssymb}
\usepackage{amsmath}
\usepackage{epstopdf}
\usepackage{gensymb}
\usepackage{natbib}
\usepackage{appendix}
\usepackage{tabu}
\usepackage{color}
\usepackage{multirow}
\usepackage{booktabs}
\usepackage{rotating}
\newcommand{\sref}[1]{Section~\ref{#1}}
\newcommand{\fref}[1]{Figure~\ref{#1}}
\newcommand{\eqn}[1]{Equation~(\ref{#1})}
\newcommand{\tab}[1]{Table~\ref{#1}}

\newcommand{\sind}{\mbox{$S$-index}}
\newcommand{\kind}{\mbox{$K$-index}}
\newcommand{\ca}{\mbox{Ca\,{\sc ii} H \& K}}

\begin{document}
\title{Modeling stellar \ca{} emission variations. I. Effect of inclination on the \texorpdfstring{$S$}{}-index.}
\accepted{March 24, 2021}

\shortauthors{Sowmya et al.}

\correspondingauthor{K.~Sowmya}
\email{krishnamurthy@mps.mpg.de}

\author[0000-0002-3243-1230]{K.~Sowmya}
\affiliation{Max-Planck-Institut f\"ur Sonnensystemforschung, Justus-von-Liebig-Weg 3, 37077 G\"ottingen, Germany}

\author[0000-0002-8842-5403]{A.~I.~Shapiro}
\affiliation{Max-Planck-Institut f\"ur Sonnensystemforschung, Justus-von-Liebig-Weg 3, 37077 G\"ottingen, Germany}

\author[0000-0002-0929-1612]{V.~Witzke}
\affiliation{Max-Planck-Institut f\"ur Sonnensystemforschung, Justus-von-Liebig-Weg 3, 37077 G\"ottingen, Germany}

\author[0000-0001-6090-1247]{N.-E.~N\`emec}
\affiliation{Max-Planck-Institut f\"ur Sonnensystemforschung, Justus-von-Liebig-Weg 3, 37077 G\"ottingen, Germany}
\affiliation{Institut f\"ur Astrophysik, Georg-August-Universit\"at G\"ottingen, Friedrich-Hund-Platz 1, 37077 G\"ottingen, Germany}

\author[0000-0002-0335-9831]{T.~Chatzistergos}
\affiliation{Max-Planck-Institut f\"ur Sonnensystemforschung, Justus-von-Liebig-Weg 3, 37077 G\"ottingen, Germany}
\affiliation{INAF Osservatorio Astronomico di Roma, Via Frascati 33, 00078 Monte Porzio Catone, Italy}

\author[0000-0003-3851-7323]{K.~L.~Yeo}
\affiliation{Max-Planck-Institut f\"ur Sonnensystemforschung, Justus-von-Liebig-Weg 3, 37077 G\"ottingen, Germany}

\author[0000-0002-1377-3067]{N.~A.~Krivova}
\affiliation{Max-Planck-Institut f\"ur Sonnensystemforschung, Justus-von-Liebig-Weg 3, 37077 G\"ottingen, Germany}

\author[0000-0002-3418-8449]{S.~K.~Solanki}
\affiliation{Max-Planck-Institut f\"ur Sonnensystemforschung, Justus-von-Liebig-Weg 3, 37077 G\"ottingen, Germany}
\affiliation{School of Space Research, Kyung Hee University, Yongin, Gyeonggi 446--701, Korea}

\begin{abstract}
The emission in the near ultraviolet \ca{} lines is modulated by stellar magnetic activity. Although this emission, quantified via the \sind{}, has been serving as a prime proxy of stellar magnetic activity for several decades, many aspects of the complex relation between stellar magnetism and \ca{} emission are still unclear. The amount of measured \ca{} emission is suspected to be affected not only by the stellar intrinsic properties but also by the inclination angle of the stellar rotation axis. Until now such an inclination effect on \sind{} has remained largely unexplored. To fill this gap, we develop a physics-based model to calculate \sind{}, focusing on the Sun. Using the distributions of solar magnetic features derived from observations together with \ca{} spectra synthesized in non-local thermodynamic equilibrium, we validate our model by successfully reconstructing the observed variations of solar \sind{} over four activity cycles. Further, using the distribution of magnetic features over the visible solar disk obtained from surface flux transport simulations, we obtain \sind{} time series dating back to 1700 and investigate the effect of inclination on \sind{} variability, both on the magnetic activity cycle and the rotational timescales. We find that when going from an equatorial to a pole-on view, the amplitude of \sind{} variations decreases weakly on the activity cycle timescale and strongly on the rotational timescale (by about 22\% and 81\%, respectively, for a cycle of intermediate strength). The absolute value of \sind{} depends only weakly on the inclination. We provide analytical expressions that model such dependencies.
\end{abstract}


\section{Introduction}
\label{sec:intro}
The magnetic field produced by the action of a dynamo in the stellar interior emerges on the stellar surface leading to the formation of magnetic features, such as dark spots and bright faculae \citep[see, e.g.,][for a review of the solar case]{Sami_B}. This magnetic field extends further into the chromosphere, where it causes a non-thermal heating resulting in the formation of plages, which are the chromospheric counterparts of faculae. This non-thermal heating of the chromosphere increases the emission in the chromospheric spectral lines, particularly in the cores of the \ca{} lines at 3968.47\,\AA{} and 3933.66\,\AA{}, respectively \citep{1970PASP...82..169L}. Consequently, observations of the \ca{} flux from a star allow monitoring its level of magnetic activity.

Since direct measurements of the \ca{} absolute flux might be affected by stability issues (e.g., spurious instrumental effects, changes in the line-of-sight air mass etc.) and do not allow comparing the magnetic activity of different stars, various indices have been introduced to overcome these issues. Their main principle is to consider ratios between the emission in the \ca{} line cores (which is strongly affected by the activity) and the emission in either the \ca{} wings or in the nearby spectral regions (which is only marginally affected by magnetic activity). One such index of the stellar magnetic activity based on the \ca{} emission is the \sind{} established by the Mt. Wilson Observatory \citep{1978PASP...90..267V,1978ApJ...226..379W}. It is defined as the ratio of the summed fluxes in the \ca{} line cores to the summed fluxes in two pseudo-continuum regions near the blue wing of the K line and red wing of the H line. Pseudo-continua are spectral regions that are closest to the true continua in a given spectral region, but are nevertheless densely populated with absorption lines.

The HK project at the Mt. Wilson Observatory (MWO HK), initiated by Olin~C.~Wilson, recorded \sind{} measurements of lower main-sequence stars since 1966. This allowed exploring stellar magnetic activity cycles \citep{1978ApJ...226..379W}. In particular, 60\%\ of the stars in Wilson's sample were found to have periodic or apparently periodic cyclic variations in activity, 25\%\ of the stars showed variations with no definite periodicity and 15\%\ displayed no significant activity variations \citep{1998ASPC..154..153B}. In a target sample consisting of 143 Sun-like stars and the Sun, monitored by a complementary synoptic survey at the Lowell observatory: the Solar Stellar Spectrograph (SSS) project, \citet{2007AJ....133..862H} found that the general distribution of cycle characteristics were in agreement with \citet{1998ASPC..154..153B}. Using the \ca{} emission data from MWO along with photometric data from Lowell and Fairborn observatories, \citet{2007ApJS..171..260L} examined the patterns of photometric and chromospheric variations among Sun-like stars (which they define as stars on or near the main-sequence having a color index $0.4\leq (B-V)\leq 1.2$). They found that variations of the chromospheric emission is correlated with the mean chromospheric activity \citep[see also][]{1998ApJS..118..239R}.

Although the HK project and the SSS survey allowed a number of breakthroughs in understanding stellar magnetic activity, they were limited to just a few hundred stars. The arrival of large spectroscopic surveys which observe hundreds of thousands of stars has rekindled the interest in stellar activity studies. For example, the Large Sky Area Multi-Object Fibre Spectroscopic Telescope (LAMOST) spectroscopic survey \citep{2012RAA....12.1197C,2012RAA....12..723Z} has collected millions of spectra in the wavelength range $3700-9100$\,\AA{}, thus providing a vast amount of \ca{} data. The significant overlap between LAMOST stars and stars observed by the {\it Kepler} space telescope \citep{2015ApJS..220...19D} allowed connecting the \ca{} emission with photometric variability. For example, \citet{2020ApJS..247....9Z} found a correlation between the photospheric activity (characterised by the level of photometric variability), chromospheric activity (characterised by the \ca{} emission), and the rotation period in late-type stars.

Despite the fact that the \sind{} is widely used as an activity indicator, its exact connection to stellar activity (defined here as coverage of the stellar surface by magnetic features) is still unclear. In particular, one can expect that the \sind{} measurements depend on the stellar inclination (defined as the angle between the stellar rotation axis and the line-of-sight of the observer). Indeed, when the inclination changes, the distribution of the magnetic features on the visible stellar disk appears differently. This, in turn, affects the contribution of the magnetic features to the measured \ca{} emission (due to the center-to-limb variations of their contrasts and foreshortening). Some studies \citep[see, e.g.,][]{2001A&A...376.1080K,2014A&A...569A..38S} employed a rather simplified modeling approach to estimate the effect of the inclination on the amplitude of the activity cycle in the \sind{}. However, a detailed investigation of the impact of the inclination on the \sind{} variations has until now been missing.

In this study, we develop a comprehensive physics-based model for calculating the \sind{}. To validate our model against observations we use the Sun which is a perfect gateway to understand other late-type stars in our galaxy and elsewhere \citep[see, e.g., a compilation of reviews in][]{best_book_ever}. We also investigate the effect of the stellar inclination on the \sind{} while the effect of the metallicity will be addressed in a separate paper. In \sref{sec:model}, we describe our model which is based on the SATIRE \citep[Spectral And Total Irradiance REconstruction;][]{2000A&A...353..380F,2003A&A...399L...1K} model. In \sref{sec:svar}, we validate our approach by comparing our model output with the available measurements of the \ca{} emission of the Sun (which obviously was observed from the ecliptic plane). In \sref{sec:incl}, we extend our approach to derive the \sind{} for other inclinations and quantify the impact of the inclination on the \sind{} on both the activity cycle and rotational timescales. In \sref{sec:concl}, we present a brief summary and some concluding remarks.

\begin{figure*}[ht!]
    \centering
    \includegraphics[scale=0.45]{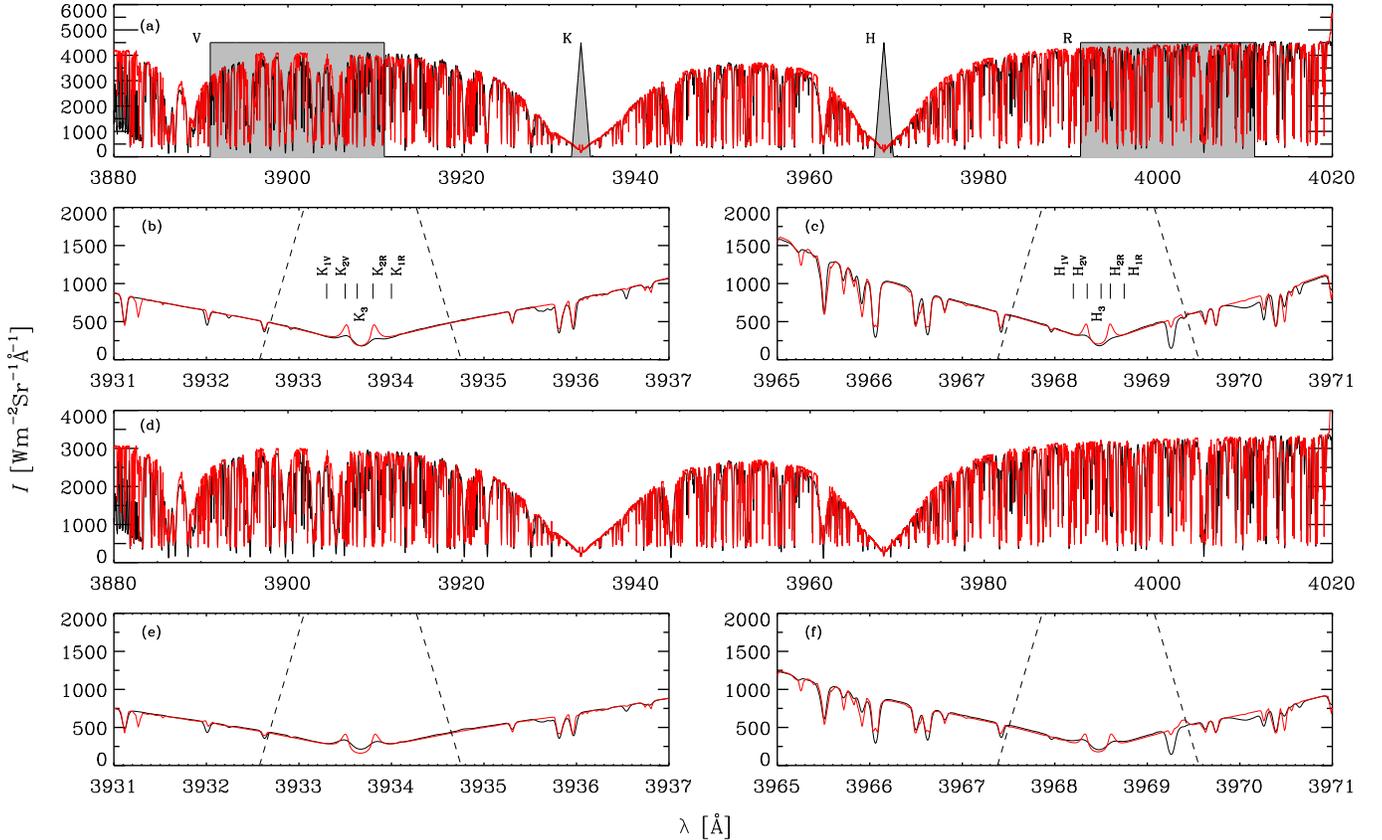}
    \caption{Comparison between the observed Fourier Transform Spectrometer (FTS) spectra (black) and the spectra synthesized with the RH code (red). Panels (a), (b), and (c) show the quiet Sun disk-center spectra while the disk-integrated quiet Sun spectra are shown in panels (d), (e), and (f). The gray shaded regions in panel (a) indicate the H, K, R and V passbands considered for the \sind{} computations. The spectral features of the K and H lines are marked in panels (b) and (c). The regions between the dashed lines in panels (b), (c), (e), and (f) highlight the parts of the spectra that fall in the K and H triangular passbands.}
    \label{fig:spectra}
\end{figure*}

\section{The model}
\label{sec:model}
\subsection{Chromospheric activity index}
\label{ssec:sind}
Following the original definition by \citet{1978PASP...90..267V}, we calculate the \sind{} as:
\begin{equation}
\centering
    S(t) =\, 8 \, \alpha_{\rm c}\, 
    \frac{N_{\rm H}(t)+N_{\rm K}(t)}{N_{\rm R}(t)+N_{\rm V}(t)}\ ,
    \label{eq:oursindex}
\end{equation}
where $t$ is the time, $N_{\rm H}(t)$ and $N_{\rm K}(t)$ are the time dependent integrated fluxes in triangular H and K bands with full-width-at-half-maximum (FWHM) of 1.09\,\AA{} centered on the cores of the \ca{} lines, respectively. $N_{\rm R}(t)$ and $N_{\rm V}(t)$ are the time dependent integrated fluxes in 20\,\AA{} wide rectangular R and V bands in the nearby pseudo-continuum region. The H, K, R and V passbands are shown as gray shaded regions in Figures~\ref{fig:spectra} and~\ref{fig:qsplage}. 

The MWO HK project is based on the measurements using two photometers, HKP-1 ($1966-1977$) and HKP-2 ($1977-2003$). Since the two instruments used slightly different R and V passbands (note that the passbands defined above and plotted in Figures~\ref{fig:spectra} and~\ref{fig:qsplage} correspond to the ones used by the HKP-2 instrument), \citet{1978PASP...90..267V} introduced a calibration factor $\alpha_{\rm c}$ for putting HKP-2 measurements on the HKP-1 scale. Using standard stars observed with both HKP-1 and HKP-2 instruments, \citet{1991ApJS...76..383D} found the value of $\alpha_{\rm c}$ to be 2.4. To make the comparison of our calculations with the observations easier, we place the \sind{} computed from our model on the same scale as HKP-2. For this it is essential to account for the difference in the duty cycles of the four channels of HKP-2 \citep{1978PASP...90..267V}. As described in \citet{1978PASP...90..267V}, the H and K channels of HKP-2 are exposed 8 times longer than the reference R and V channels. Consequently, the H and K fluxes in \eqn{eq:oursindex} are scaled up by a factor of 8 \citep[see, e.g.,][]{2007AJ....133..862H}.

\subsection{SATIRE}
\label{ssec:sat}
We base our model on SATIRE \citep{2000A&A...353..380F,2003A&A...399L...1K} which is one of the most advanced physics-based models of solar irradiance variability. In particular, it has been validated against numerous observational records of solar irradiance \citep[see][and references therein]{balletal2014,2014A&A...570A..85Y,Sanja2016}. In SATIRE, the visible solar disk is decomposed into four components: faculae ($f$), sunspot umbrae ($u$), sunspot penumbrae ($p$), and quiet Sun ($q$, i.e. the regions of the solar disk which are not occupied by any of the magnetic features). The emergent spectra for each of these components are assumed to be time-invariant, but depend on the position on the disk. The full-disk solar spectrum is calculated by weighting the spectra of the individual components with the corresponding disk area coverages and summing up the contributions of the individual components. The distribution of the magnetic features on the solar disk and, consequently, their disk area coverages change with time, leading to the variability of the full-disk solar spectrum. 

Traditionally, SATIRE employs the spectra for the quiet Sun and magnetic features from \citet{1999A&A...345..635U}, synthesized using the ATLAS9 code \citep{1992RMxAA..23...45K,1994A&A...281..817C} under the assumption of local thermodynamic equilibrium \citep[LTE, see, however, ][for the first non-LTE version of SATIRE]{Rinat_2019}. Since the \ca{} line core emissions form in the chromosphere, where the LTE approximation is no longer valid, we replace the emergent LTE spectra from ATLAS9 with the non-LTE spectra synthesized using the RH code \citep{2001ApJ...557..389U}. A description of the spectral synthesis using the RH code is given in \sref{ssec:rh}.

Calculation of the solar surface coverages by magnetic features is another important part of SATIRE, and different approaches are taken depending on the branch of the model \citep{krivova2011}. In this study we use different branches of SATIRE to achieve two main goals. Firstly, to validate our approach, we synthesize a solar \sind{} time series and compare it to the available data (see \sref{sec:svar}). For this, we employ the \mbox{SATIRE-S} branch, which uses direct measurements of the solar surface magnetic field and continuum intensity images to derive the surface coverages by magnetic features \citep{2014A&A...570A..85Y}. Our second goal is to model the \sind{} time series back to the year 1700 and to study the dependence of \sind{} on the inclination (see \sref{sec:incl}) on both the activity cycle and the rotational timescales. Since observations of the Sun from outside the ecliptic plane are currently unavailable, we have to rely on numerical simulations to achieve this goal. Therefore, we use the surface area coverages by magnetic features calculated by \citet{Nina2, Nina1} following an approach similar to \mbox{SATIRE-T2} \citep{Dasi2016}. Further details are given in \sref{sec:incl}.

To compute the disk integrated flux at a given time $t$, and a given wavelength $\lambda$, we divide the solar disk into `$l$' concentric rings. The position of each ring is given by $\mu_l$ (defined as the cosine of the heliocentric angle $\theta$). The disk integrated spectral flux is then the sum of fluxes from each of these rings:
\begin{equation}
\begin{split}
    F(t,\lambda) & = \sum_{l} \bigg\{I_q(\lambda,\mu_l)+\\& \sum_{j}A_{\rm exp}\alpha_{jl}(t)\left[I_j(\lambda,\mu_l)-I_q(\lambda,\mu_l)\right]\bigg\}\ \Delta\Omega_l\ .
\end{split}
    \label{eq:ssi}
\end{equation}
Here, the index $j$ designates the different magnetic features, $j=\{f,u,p\}$. $I_q$, $I_f, I_u, I_p$ are the intensities emerging from the quiet Sun, faculae, umbrae, and penumbrae, respectively. $\alpha_{jl}(t)$ are the inclination-dependent fractional coverages of the visible solar disk by a given magnetic component in the $l^{\rm th}$ ring. $\Delta\Omega_l$ is the solid angle subtended by the $l^{\rm th}$ ring to an observer located at the distance of 1 AU from the Sun. The term inside the square brackets represents the intensity contrasts of the magnetic features with respect to the quiet Sun. 

We note that the $\alpha_{jl}(t)$ values used in our model are deduced from photospheric measurements (in \mbox{SATIRE-S}) and simulations (in \citealt{Nina1}) and hence correspond to area coverages in the low/mid photosphere. At the same time, most of the \ca{} radiation in passbands having a typical width of about 1\,\AA{} comes from the photosphere below 550\,km \citep[see e.g. Figure~3 of][showing the contribution function for the Ca\,{\sc ii} radiation transmitted through the SUNRISE/SuFI Ca\,{\sc ii} H filter \citealt{2010ApJ...723L.127S,2011SoPh..268...35G}]{2017ApJS..229...11J}, with a secondary peak in the contribution in the chromosphere. The center-of-gravity of the contribution functions lies between 500 and 700\,km. The flux tubes forming the magnetic structures expand as they rise to these heights to maintain horizontal pressure balance \citep{1990A&A...234..519S,Sami_B}. Such an expansion means that the magnetic features cover a larger fraction of the solar disk at these heights \citep[e.g.][]{1991A&A...250..220S}. To account for this expansion, we multiply the photospheric area coverages by a constant factor $A_{\rm exp}$ (see \sref{ssec:calib} for more details).

We obtain the integrated fluxes in the H, K, R, and V bands using
\begin{equation}
    N_m(t) = \int_m F(t,\lambda_m)\ T(\lambda_m)\ \ d\lambda_m\ ,
    \label{eq:intflux}
\end{equation}
where $m=\{{\rm H},{\rm K},{\rm R},{\rm V}\}$ correspond to the four passbands. The wavelength integration is performed over the interval defined by $\lambda_m$ for each of these bands, after multiplying the spectral flux in that interval with the transmission profile of the corresponding passband defined by $T(\lambda_m)$.
\begin{figure*}[ht!]
    \centering
    \includegraphics[scale=0.5]{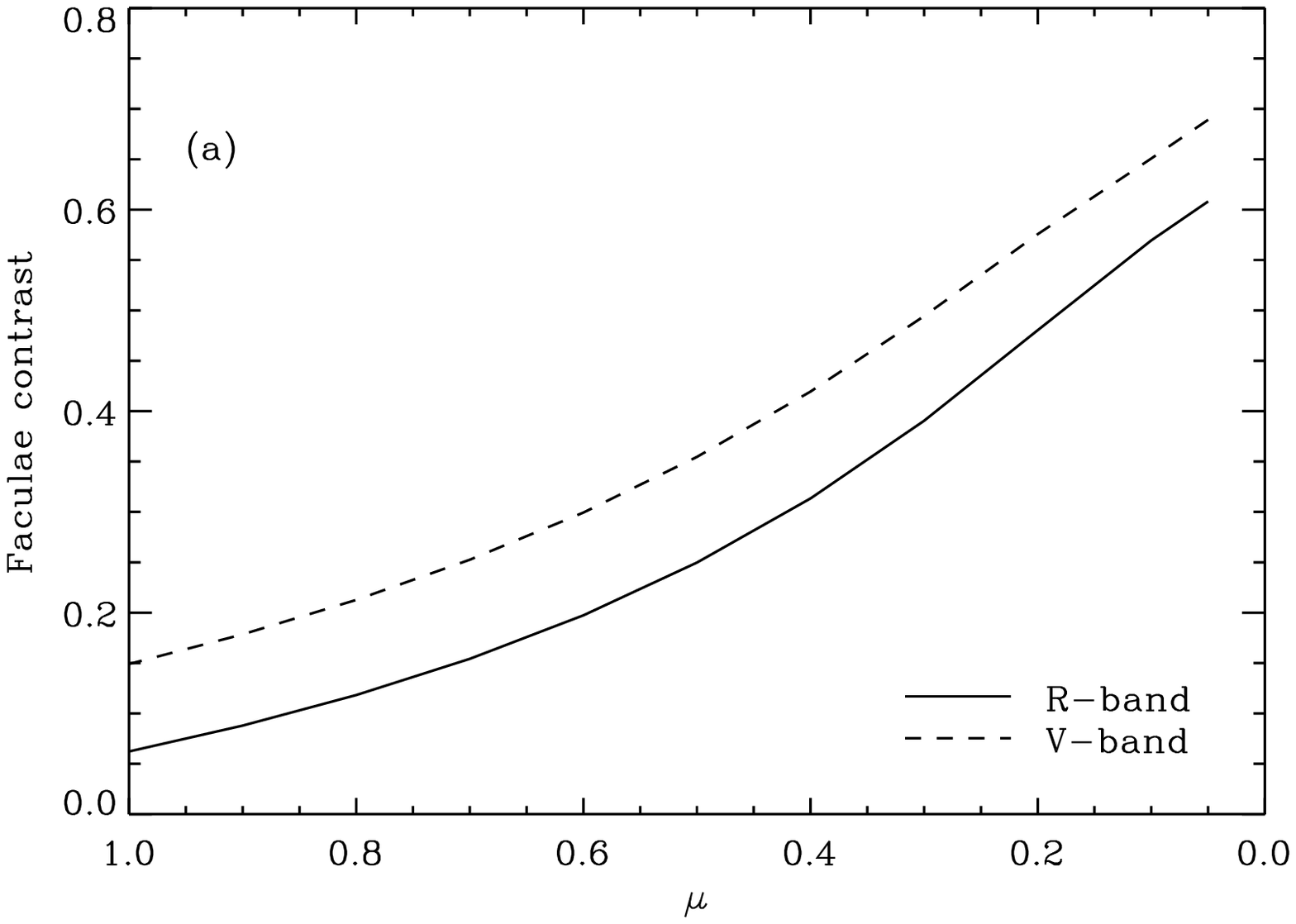}
    \includegraphics[scale=0.5]{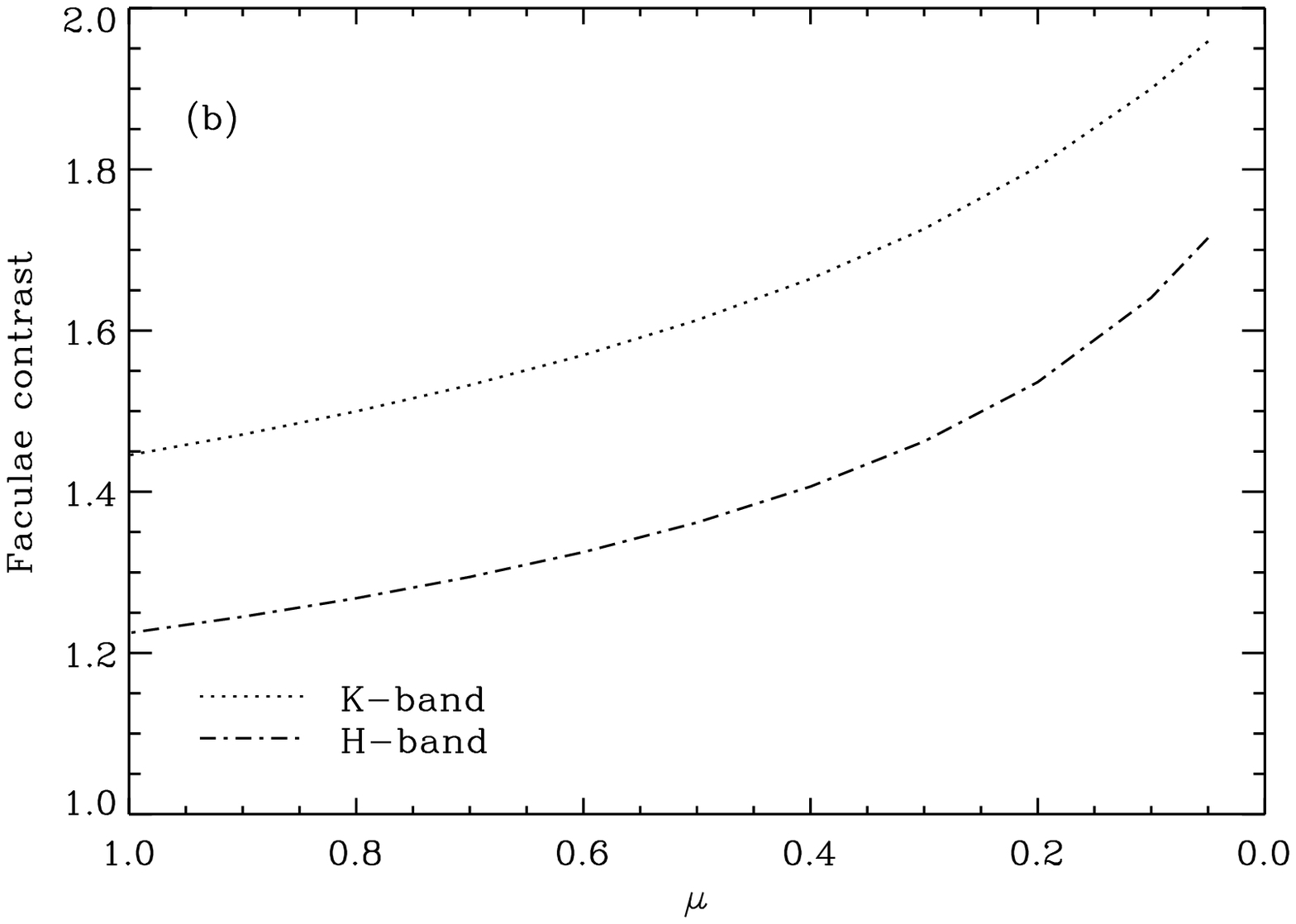}
    \caption{Center-to-limb variation of the intensity contrast of the faculae with respect to the quiet Sun, defined as $\{I_f(\lambda,\mu_l)-I_q(\lambda,\mu_l)\}/I_q(\lambda,\mu_l)$, in the pseudo-continuum R and V passbands (panel a) and in the line core H and K passbands (panel b).}
    \label{fig:contrast}
\end{figure*}

\subsection{Spectral synthesis}
\label{ssec:rh}
When synthesizing the H and K resonance lines of singly ionized calcium, a few aspects of their formation in the solar atmosphere need to be considered. For instance, their formation spans a wide range of heights in the solar atmosphere \citep{1981ApJS...45..635V}. The \ca{} wings form in the photosphere under LTE, whereas the cores form in the chromosphere, where the densities are so low that departures from LTE become significant. Synthesizing line cores under the approximation of LTE gives rise to huge unrealistic emissions owing to an increase in the temperature of the layers above the temperature minimum. Also, partial frequency redistribution (PRD) effects are important for their formation \citep{1990ASPC....9..103U}. Therefore, a proper treatment of non-LTE and PRD effects is crucial to correctly model the \ca{} line profiles. 

We use the numerical radiative transfer code RH \citep{2001ApJ...557..389U} to synthesize both the \ca{} lines and the pseudo-continuum for different positions ($\mu_l$) on the solar disk. This code is based on the multi-level approximate lambda iteration formalism of \citet{1991A&A...245..171R,1992A&A...262..209R}. It solves the equations of statistical equilibrium and radiative transfer self-consistently for multi-level atoms and molecules. The non-LTE and PRD capability of RH makes it well suited for calculating \ca{} emission. 

The atomic model that we use for the Ca\,{\sc ii} calculations consists of five levels plus continuum. The \ca{} lines at 3968.47\,\AA{} and 3933.66\,\AA{}, respectively, and the infrared triplet transitions at 8542\,\AA{} occur between these five levels. Calcium is treated in non-LTE taking into account the effects of PRD in the H and K lines, while the transitions in other background elements are treated in LTE. The line opacity and emissivity in the wavelength range $3880-4020$\,\AA{} (see \fref{fig:spectra}a) are calculated by including 7216 transitions tabulated in Kurucz's line list\footnote{\url{http://kurucz.harvard.edu/linelists.html}}, assuming LTE. The elemental solar abundances used are from \citet{1989GeCoA..53..197A}, to remain consistent with the SATIRE model \citep[see e.g.][]{1999A&A...345..635U}.

We used the one-dimensional (1D) semi-empirical models FALC99 and FALP99 by \citet{1999ApJ...518..480F} to synthesize the spectra of the quiet Sun and faculae, respectively. We note that the contributions from sunspot umbra and penumbra to the \sind{} are ignored, because (i) we assume that sunspots are dark in both the \ca{} lines and the pseudo-continuum and thus their contributions affect the denominator and the numerator of \eqn{eq:oursindex} equally \citep[so that the \sind{} is unaffected, see also][for a more detailed discussion]{2014A&A...569A..38S}; (ii) the solar disk coverage by faculae is significantly larger than that by spots; (iii) the existing 1D semi-empirical models of umbra and penumbra lack well constrained chromospheres \citep{2017ApJ...850...35L} needed for the proper treatment of line cores. Therefore, in our model, the variations in the \sind{} are driven entirely by the faculae on the solar disk.

When computing the spectra using these models, we run into a situation where the absolute flux in the pseudo-continuum region around \ca{} lines is overestimated for the FALC99 and FALP99 models. This is a known effect which is due to the fact that the existing line lists are not complete and underestimate the opacities in the ultraviolet \citep[UV; see, e.g.,][]{2009AIPC.1171...43K}. To account for the missing opacity and to match the observed flux in the near-UV region, we used opacity fudge factors \citep{1992A&A...265..237B,2019SoPh..294..165R}. These factors are wavelength dependent, and the continuum opacity is multiplied by them to compensate for the missing atomic and molecular lines and resulting overestimation of the UV flux. The opacity fudging in our approach is done by increasing the H$^-$ opacity in the wavelength range $3700-4150$\,\AA{}, starting with the best fit multipliers used in \citet{1992A&A...265..237B}. In order to match the observed flux, we require opacity fudge factors of 1.45 at 3700\,\AA{}, 1.3 at 3900\,\AA{} and 1 (original opacity values) at 4150\,\AA{}. The fudge factors in the wavelength range $3900-4150$\,\AA{} are linearly interpolated between these values. The values of the opacity fudge factors that we used are only slightly different from those of \citet{1992A&A...265..237B}, which are respectively 1.35, 1.21 and 1.035 at the wavelengths mentioned.

To test the quiet Sun spectra synthesized using the RH code with the opacity fudging as described above, we used the high-resolution disk-center and disk-integrated atlases from the Hamburg Observatory\footnote{\url{ftp://ftp.hs.uni-hamburg.de/pub/outgoing/FTS-Atlas/}} \citep{1999SoPh..184..421N,2016A&A...590A.118D}. They are based on the data from the Fourier Transform Spectrometer (FTS) mounted on the McMath-Pierce solar telescope at the Kitt Peak National Observatory in Arizona, United States \citep[see, e.g.,][]{1984SoPh...90..205N}. We refer to the spectra from these atlases as the FTS spectra. Figure~\ref{fig:spectra} shows a comparison between the FTS spectra (in black) and the spectra synthesized with the RH code using the FALC99 model atmosphere (in red), for the disk-center and disk-integrated cases (see also \fref{fig:qsplage} for a comparison between the quiet Sun and the facular spectra).

The agreement between the pseudo-continuum in the FTS and RH spectra for both the disk-center and the disk-integrated cases is quite good. The broad \ca{} lines are well reproduced except for the K$_2$ and H$_2$ features, as shown in Figs.~\ref{fig:spectra}b and c (see also Figs.~\ref{fig:spectra}e and f). Contrary to what is observed, the K$_{\rm 2V}$ and K$_{\rm 2R}$ (similarly the H$_{\rm 2V}$ and H$_{\rm 2R}$) peaks in the synthetic spectra are symmetric and show enhanced emission. This is a general characteristic of the 1D models and is partly because the velocity gradients in the solar atmosphere are not included in our computations using 1D models. An example of what happens to the K$_2$ and H$_2$ features when macroturbulent velocity fields in the solar atmosphere are taken into account in a simple manner is shown in \fref{fig:macro}. A proper treatment of the velocity field gradients using 3D model atmospheres along with the spatial averaging of the intensity profiles is expected to make the K$_2$ and H$_2$ features asymmetric in addition to decreasing the emission strength \citep[see, e.g.,][]{2018A&A...611A..62B}. Using 3D atmospheres \citet{2018A&A...611A..62B} attempted to model the \ca{} lines. Though they could obtain asymmetric central peaks, the synthesized profiles did not yet fully reproduce the observed disk-center profiles indicating that further effort in improving 3D models is needed.

To evaluate the extent to which the \sind{} values are affected by the slight mismatch between the FTS and RH spectra, we computed the flux ratio in \eqn{eq:oursindex}. Namely, we employed the disk-center and disk-integrated spectra from FTS and their RH counterparts computed with the FALC99 model (i.e. the disk-center spectra shown in Figs.~\ref{fig:spectra}a--c and the full-disk spectra shown in Figs.~\ref{fig:spectra}d--f) to compute  $S_{\rm RH}$ and $S_{\rm FTS}$. For the disk-center spectra the ratio $S_{\rm RH}$/$S_{\rm FTS}$ is found to be 1.04, while for the disk-integrated spectra it is equal to 0.95. We note that while the FTS disk-center spectra are recorded by choosing regions on the solar disk that are free of any apparent magnetic activity, the disk-integrated spectra are bound to be affected by the magnetic activity. Therefore, it is not surprising that the ratio is smaller for the disk-integrated spectra. Most importantly, both ratios are very close to unity so that the mismatch between the observed and synthesized spectra is not expected to noticeably affect our calculations.

Finally, we note that the center-to-limb variation (CLV) of the facular contrast in the line core and pseudo-continuum passbands is one of the key factors defining the \sind{} dependence on the inclination. Observations of faculae at continuum wavelengths show that their contrast relative to the quiet Sun is much lower at the disk center than at the limb \citep{1992ApJ...396..351T}. This is because the hot walls of the small flux tubes forming faculae are hardly visible closer to the disk center and become best visible at the limb \citep{1976SoPh...50..269S}. Naturally, the 1D models cannot directly account for such a purely 3D effect. At the same time, the temperature structure of the semi-empirical FALP99 model has been created to emulate this effect and has been demonstrated to reproduce available measurements of the facular contrast rather well \citep[see ][for a detailed discussion]{1999ApJ...518..480F}. Figure~\ref{fig:contrast} shows the calculated CLV of facular contrast defined as, [$\{I_f(\lambda,\mu_l)-I_q(\lambda,\mu_l)\}/I_q(\lambda,\mu_l)$], where $I_f$ and $I_q$ are the faculae and quiet-Sun intensities computed using the FALP99 and FALC99 models, respectively. Both the pseudo-continuum and line core contrasts increase monotonously towards the limb (see Figs.~\ref{fig:contrast}a and b, respectively). However, the magnitude of this increase in the pseudo-continuum passbands is significantly different from that in the line core passbands. The CLV of the contrast is very steep in the pseudo-continuum, i.e. the contrast at the limb increases to nearly 10 times the value at the disk center, whereas in the line cores the increase is much less pronounced (limb contrast is only 1.5 times the value at $\mu=1$). Also, the line core contrasts are overall higher due to additional non-thermal heating in the higher layers of the solar atmosphere where they form. In \sref{sec:incl} we show how these properties of the CLV define the dependence of the \sind{} on the inclination.

\section{Solar \texorpdfstring{$S$}{}-index variations}
\label{sec:svar}
\subsection{Calibration of the model using solar observations}
\label{ssec:calib}
The MWO HK photometers have made direct measurements of the \sind{} of the Sun using the sunlight reflected from the Moon \citep{2017ApJ...835...25E}. Other records of \sind{} proxies include the \kind{} (defined as the equivalent width of a 1\,\AA{} band centered on the K line) data from the National Solar Observatory Kitt Peak \citep[NSO/KP;][]{1978ApJ...226..679W}, Sacramento Peak \citep[NSO/SP;][]{1984ApJ...276..766K}, and the Synoptic Optical Long-term Investigations of the Sun by the Integrated Sunlight Spectrometer \citep[SOLIS/ISS;][]{2011SPIE.8148E..09B} at Kitt Peak. The NSO/KP measurements cover the time period $1974-2016$, NSO/SP the period $1976-2016$, and SOLIS/ISS between 2006 and present, thus complementing one another.
\begin{figure}[ht!]
    \centering
    \includegraphics[scale = 0.495,trim=0.5cm 0.5cm 0.1cm 4.5cm,clip]{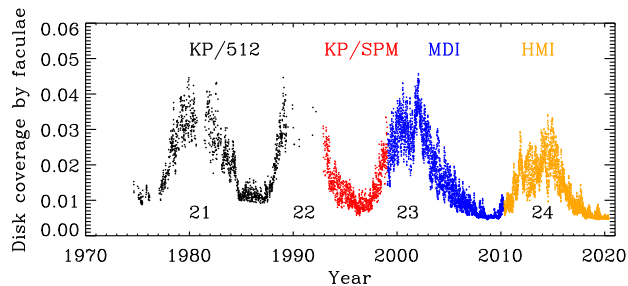}
    \caption{Disk area coverage by the facular component on the photosphere for solar cycles $21-24$ obtained using the approach of \citet{2014A&A...570A..85Y}. The colors correspond to observations from different instruments as indicated.}
    \label{fig:ffactors}
\end{figure}
\begin{figure*}[ht!]
    \centering
    \includegraphics[scale=0.395]{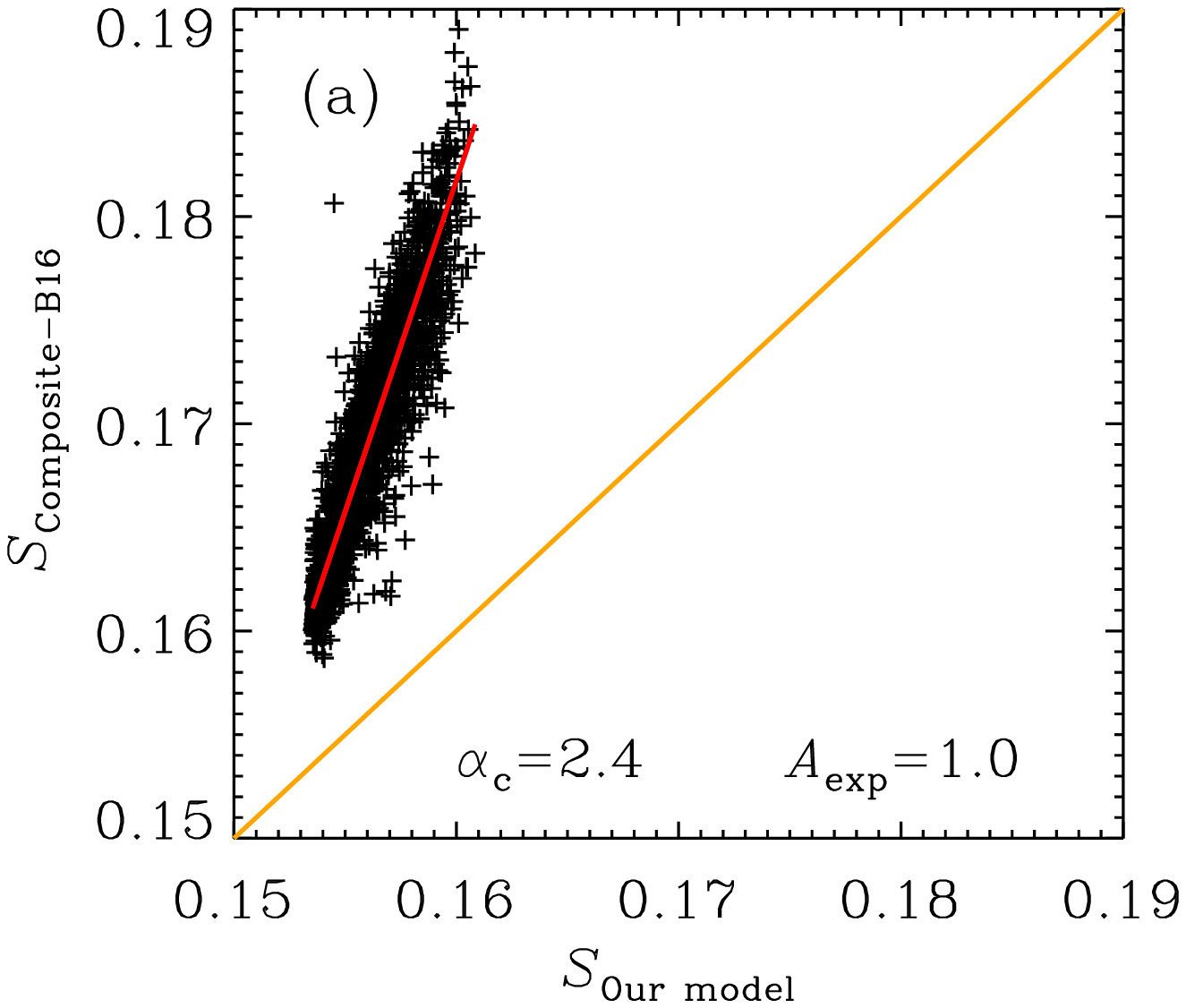}
    \includegraphics[scale=0.395]{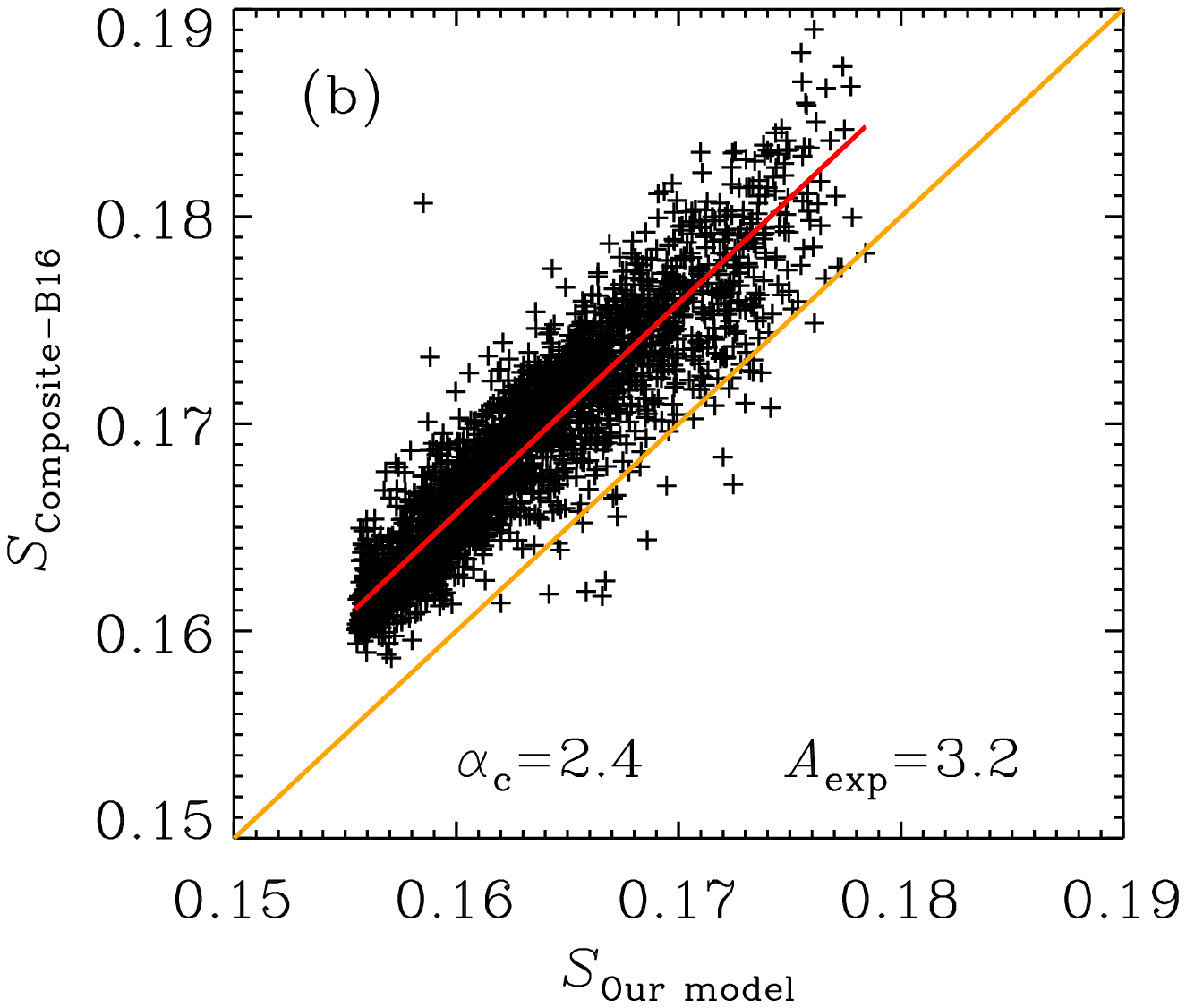}
    \includegraphics[scale=0.395]{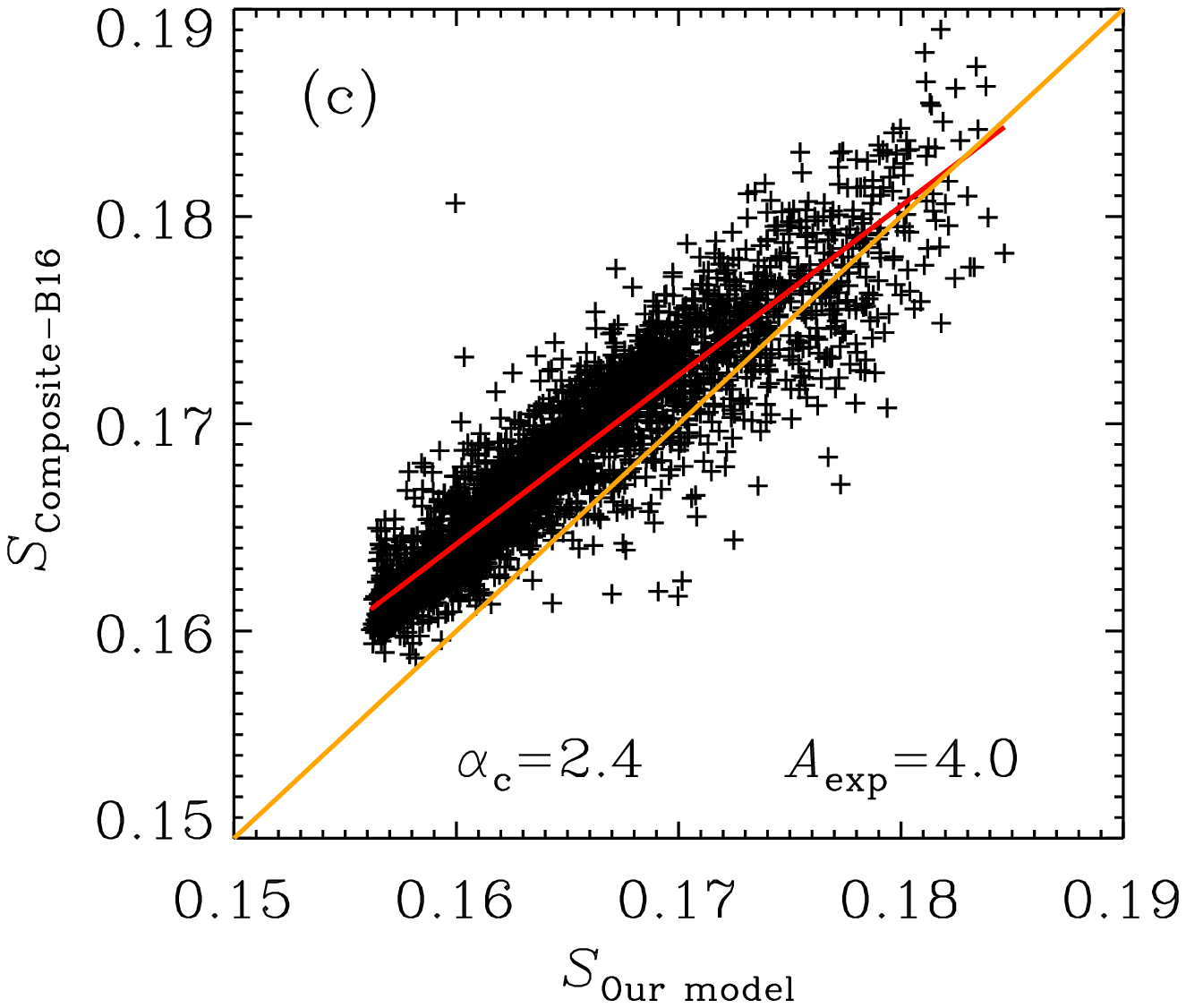}
    \includegraphics[scale=0.395]{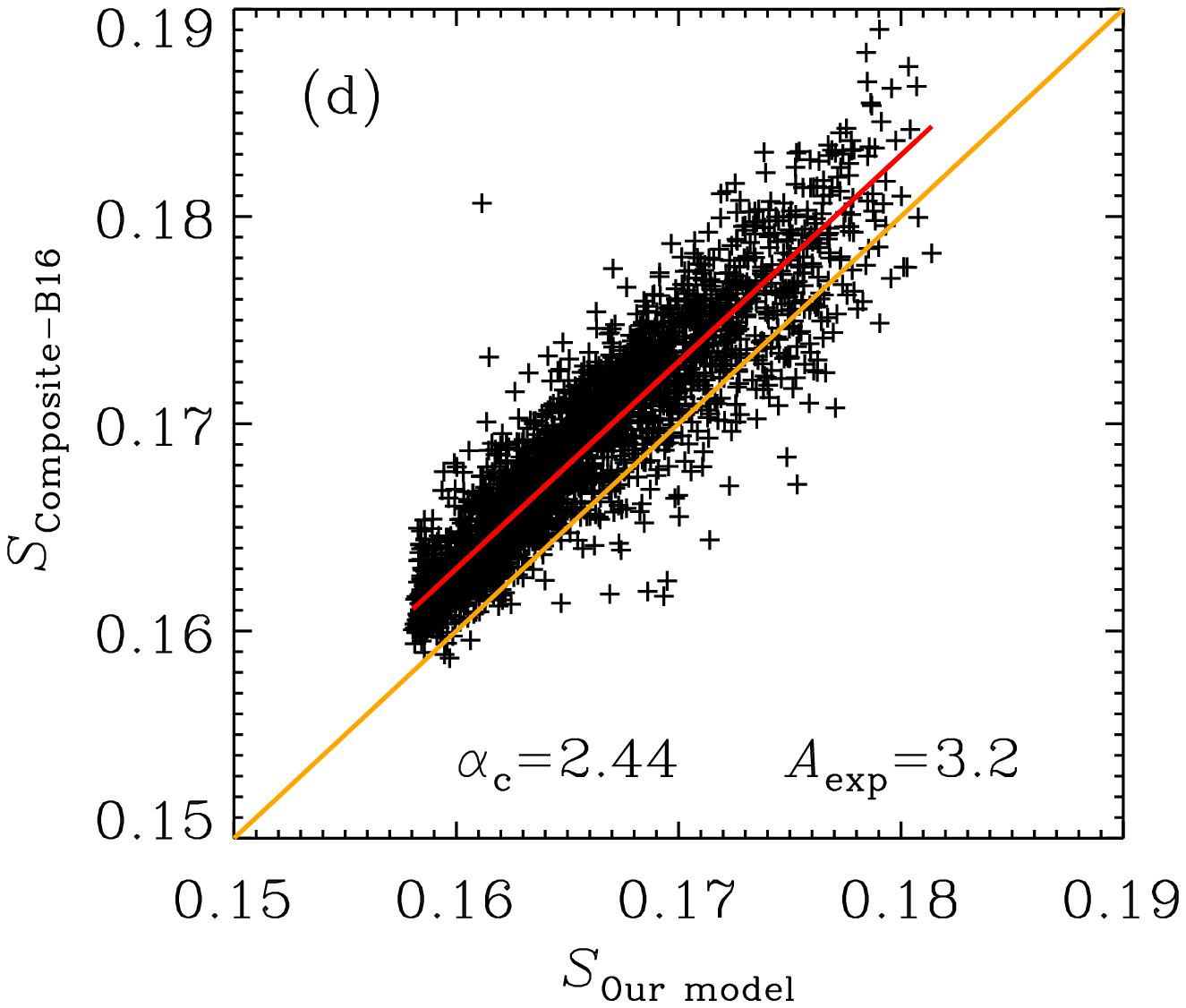}
    \includegraphics[scale=0.395]{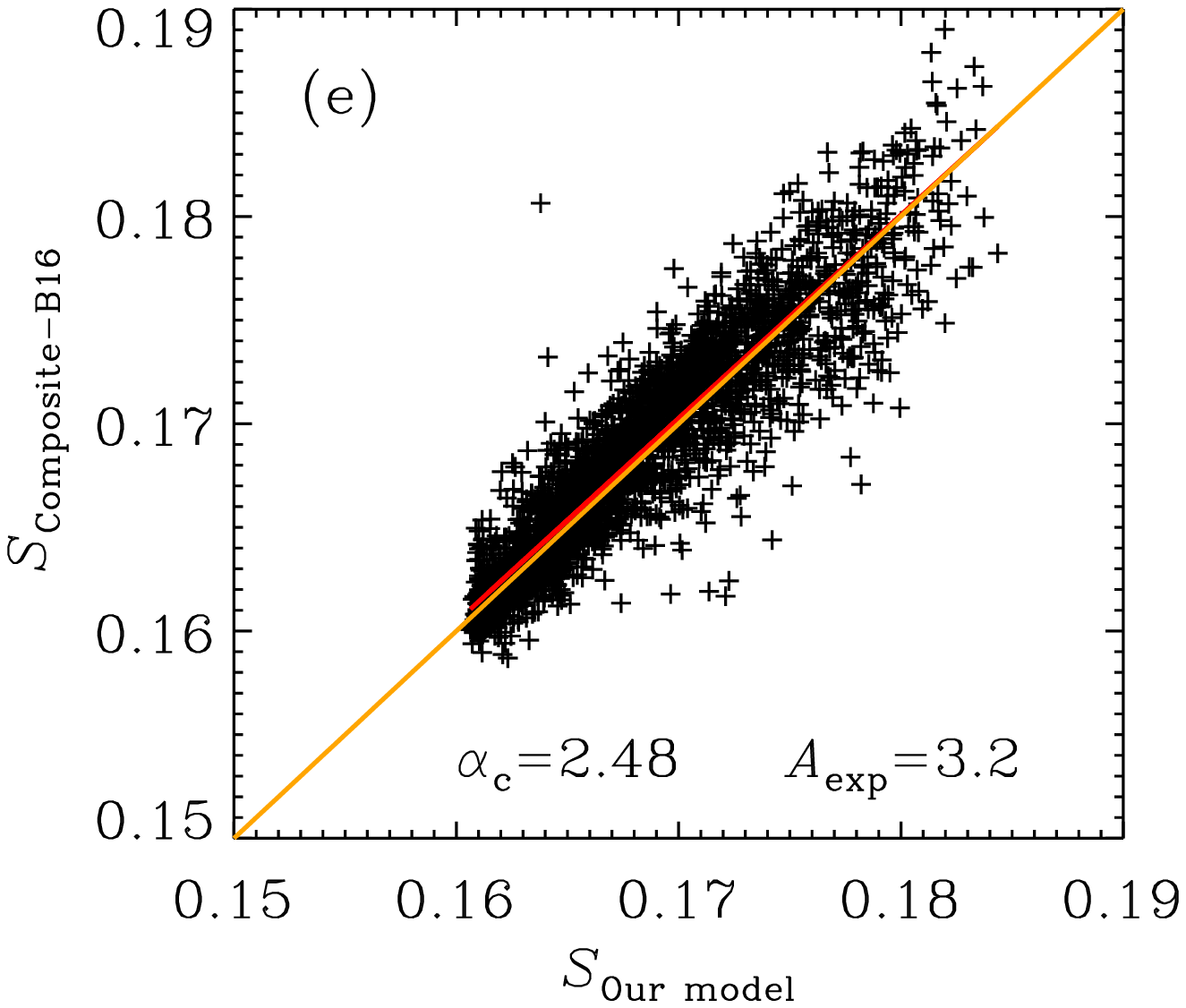}
    \includegraphics[scale=0.395]{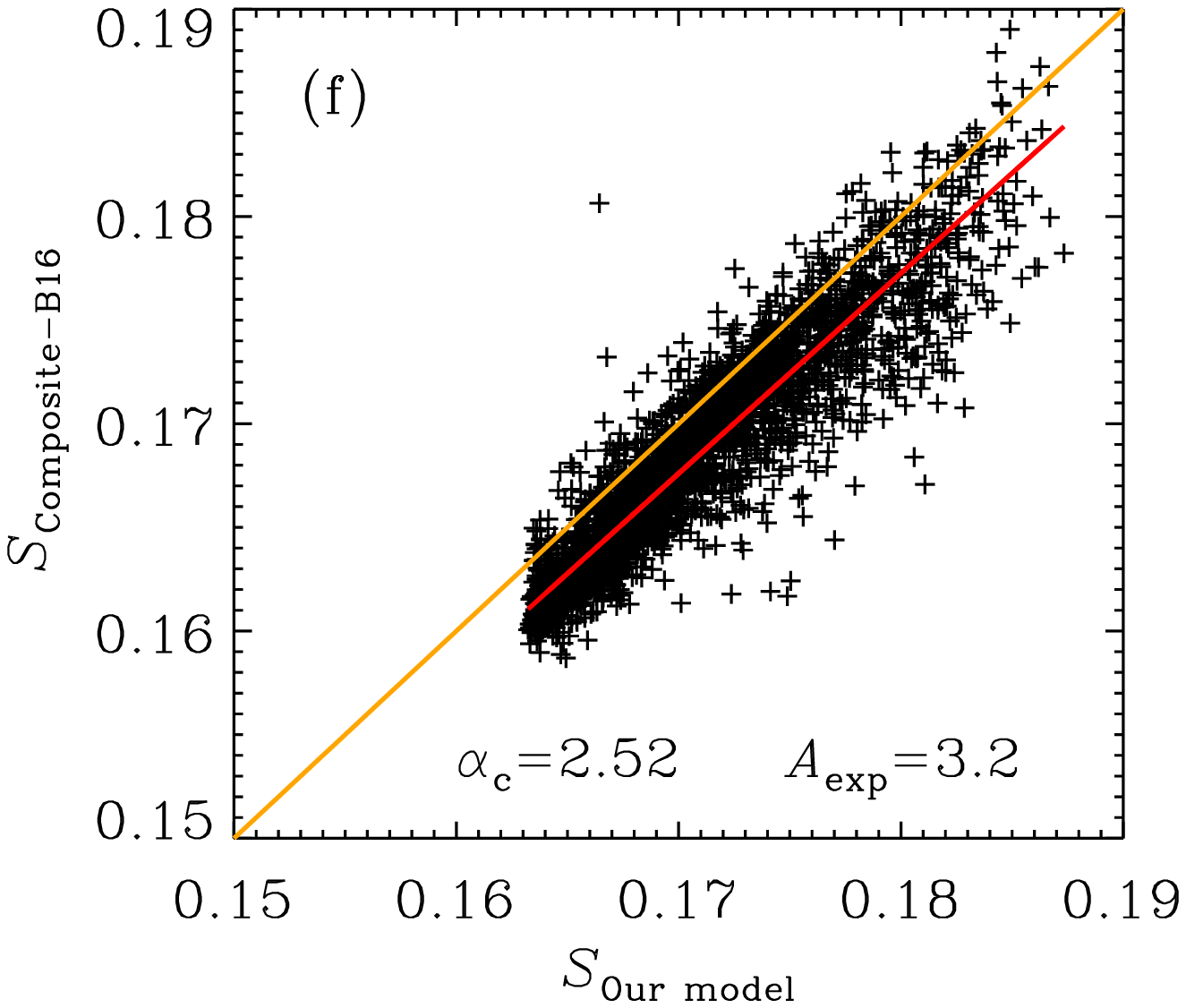}
    \caption{Observed (\mbox{Composite-B16}) vs. calculated values of the \sind{} for different combinations of the parameters $\alpha_{\rm c}$ and $A_{\rm exp}$ as indicated in each panel. The red line in each panel shows the linear fit to the data and the orange line indicates the expectation value. The value of the linear Pearson correlation coefficient is 0.93.}
    \label{fig:lrbersat}
\end{figure*}

By establishing a linear relationship between the NSO/SP \kind{} and the MWO \sind{}, \citet{2017ApJ...835...25E,2017yCat..18350025E} derived and published a composite time series of the daily \sind{} values calibrated to the MWO/HKP-2 scale for the solar cycles $20-24$. We call this composite as `\mbox{Composite-E17}'. Taking the advantage of the data from MWO/HKP-2, \citet{2017ApJ...835...25E} also accurately placed solar cycle 23 on the \sind{} scale of HKP-2. \citet{2016SoPh..291.2967B} established a relation between the NSO/SP and SOLIS/ISS measurements and created a composite of the daily \kind{} values on the SOLIS/ISS scale\footnote{\url{https://solis.nso.edu/0/iss/}, see also Technical Report No. NSO/NISP-2007-001}, which we call $K_{\rm ISS}$. We use the \kind{} composite from \citet{2016SoPh..291.2967B} and transform it to the \sind{} on the MWO/HKP-2 scale using the following linear relation from \citet{2017ApJ...835...25E}:
\begin{equation}
    S(K_{\rm ISS}) = 1.71\ K_{\rm ISS}\ +\ 0.02\ .
    \label{eq:skiss}
\end{equation}
We refer to the \sind{} composite obtained using \eqn{eq:skiss} as `\mbox{Composite-B16}'.

We use the \mbox{Composite-B16} to determine the value of the expansion factor $A_{\rm exp}$ (introduced in \sref{ssec:sat}) and to test our calculations. The photospheric area coverages by faculae needed for this purpose are obtained from \mbox{SATIRE-S} \citep{2014A&A...570A..85Y} for the period covering solar cycles $21-24$. Figure~\ref{fig:ffactors} shows the fraction of the solar disk covered by faculae derived from the NSO Kitt Peak 512-channel Diode Array Magnetograph (black dots) and the spectromagnetograph (SPM; red dots), as well as the Michelson Doppler Imager (MDI; blue dots) onboard the Solar and Heliospheric Observatory, and the Helioseismic and Magnetic Imager (HMI; orange dots) onboard the Solar Dynamics Observatory. The gaps in the data are due to the limited availability of magnetograms suitable for calculating the area fractions.

Next, by employing the synthetic spectra from RH in combination with the facular area coverages i.e. the area fraction covered by magnetic flux giving rise to faculae shown in \fref{fig:ffactors}, we computed the \sind{} using \eqn{eq:oursindex}. We then regressed the daily \sind{} values from the \mbox{Composite-B16} to those returned by our calculations (excluding the data gaps i.e. keeping only the common days). The result is shown in \fref{fig:lrbersat}a. The red line gives the linear fit to the data and the orange line shows the expectation values for a perfect match. The linear Pearson correlation coefficient between the model and the measurements is $r=0.93$. Such a high correlation is in itself reassuring. However, the computed \sind{} values were much lower than the observed ones. To bring the modeled values closer to the expectation, we varied two parameters. One is the calibration constant $\alpha_{\rm c}$ introduced in \sref{ssec:sind} which affects the overall level of the calculated \sind{} and the other is the active area expansion factor $A_{\rm exp}$ which governs the cycle amplitudes as well as the amplitudes of the variability on all timescales. Panels a--c and d--f in \fref{fig:lrbersat} show how the slope of the regression changes with $A_{\rm exp}$ and the offset with  $\alpha_{\rm c}$, respectively. By visual inspection, we chose $A_{\rm exp}=3.2$ and $\alpha_{\rm c}=2.48$ since these values led to a linear fit basically coinciding with the expectation value (panel e).

We note that this value of $A_{\rm exp}$ is consistent with the results by \cite{chatzistergos_analysis_2017}, where an approximate value of 3 is suggested when comparing the plage areas derived from full-disk Ca\,{\sc ii} K observations to the facular areas obtained from HMI magnetograms. Here, we have assumed that the expansion factor does not depend on the size of the flux tube. We believe that this is a reasonable assumption since \citet{1999A&A...347L..27S} have shown that in the photosphere, flux tubes of different sizes expand at nearly the same relative rate implying that the expansion factor is roughly constant. According to Figure~4 of \citet{1986A&A...154..231P}, the area covered by a flux tube with magnetic filling factor about 10\%\ at a height of 500\,km (where the Ca\,{\sc ii} radiation in the H and K passbands come from) is about 4 times that at 250\,km (where the spectral lines typically used for magnetogram measurements form). This area ratio is constant for all magnetic filling factors of 10\%\ and below, i.e. for network and moderate plage. For strong plage, e.g., with a magnetic filling factor of 25\%, the area ratio is about 2. Therefore, for a mix of filling factors (both larger and smaller than 10\%), the value of 3.2 for the expansion factor looks reasonable.
 
The deviation of $\alpha_{\rm c}$ from its original value of 2.4 in \citet{1991ApJS...76..383D} by a modest 3.3\%\ is not entirely surprising. It is partly because the observed features of the \ca{} lines are not perfectly reproduced in the full-disk spectra synthesized using RH. It can also be partly due to the inaccuracies associated with the triangular passbands i.e. the triangular functions that we use for the H and K passbands are probably not fully representative of the transmission profiles of the H and K channels of MWO/HKP-2.

\subsection{Reconstruction of the solar \texorpdfstring{$S$}{}-index}
\label{ssec:rec}
\begin{figure*}[ht!]
    \centering
    \includegraphics[scale=0.9,trim=0.cm 0.5cm 0.cm 0.cm,clip]{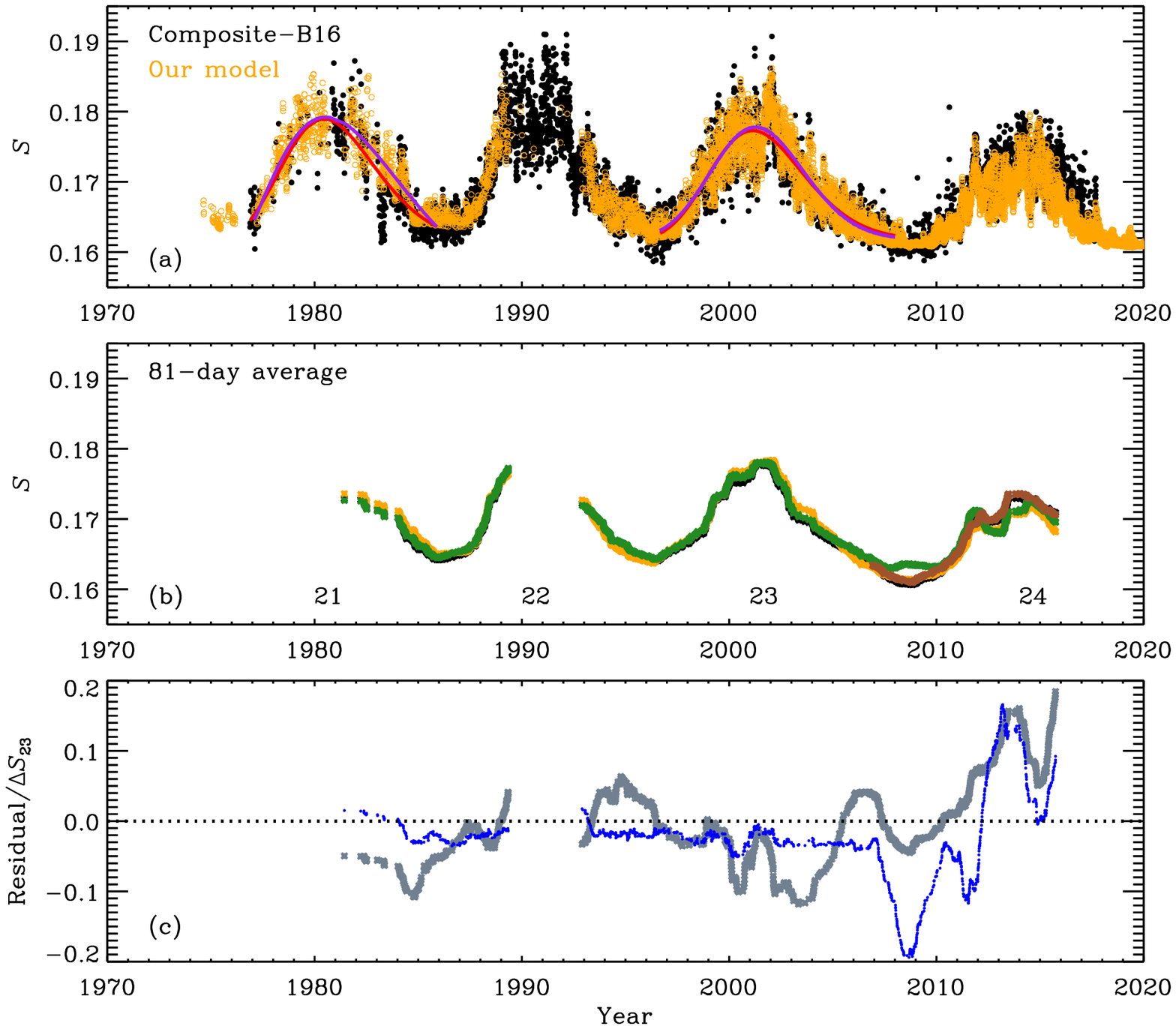}
    \caption{Panel (a): comparison between the measured \sind{} from the \mbox{Composite-B16} (black circles) and the \sind{} reconstructed with our model (orange circles). The red curves are the cycle shape fits to the \mbox{Composite-B16} and the purple curves are the fits to our model calculations. Panel (b): 81-day averages of the same day values from \mbox{Composite-B16} (black curve), NSO/SP (green) and our calculations (orange curve). For comparison, SOLIS/ISS values (brown) are shown. The solar cycle numbers are indicated below the curves. Panel (c): difference in the 81-day smoothed values between \mbox{Composite-B16} and our model (gray), and between \mbox{Composite-B16} and NSO/SP (blue), normalized to the cycle 23 amplitude for \mbox{Composite-B16} ($=0.0149$; see \tab{tab:sparams}).}
    \label{fig:sindexbersatnew}
\end{figure*}

Figure~\ref{fig:sindexbersatnew}a compares the daily values of the \sind{} returned by our model to the \mbox{Composite-B16}. The 81-day moving averages of the daily values are shown in \fref{fig:sindexbersatnew}b. Here, we show not only \mbox{Composite-B16} but also its components, namely, NSO/SP and SOLIS/ISS data. For the 81-day averages, only those days are used for which data from all three sources: \mbox{Composite-B16} (black), NSO/SP (green) and our model (orange) were available. The NSO/SP \sind{} values are extracted from \mbox{Composite-E17}. For comparison, the data from SOLIS/ISS (brown) are also shown. The difference between \mbox{Composite-B16} and our model values in \fref{fig:sindexbersatnew}b (i.e. the residual) is shown in gray in \fref{fig:sindexbersatnew}c. The residuals have been normalized to the amplitude of solar cycle 23 for \mbox{Composite-B16} given in \tab{tab:sparams}. The residuals lie within 20\%\ of the cycle 23 amplitude, showing that the reconstructed \sind{} agrees reasonably well with the observed one (not surprising, given the good match in \fref{fig:lrbersat}e). We remind the reader that our model was calibrated to match \mbox{Composite-B16} using two free parameters $\alpha_{\rm c}$ and $A_{\rm exp}$. The larger uncertainties for solar cycle 24 could be because for this cycle, the \mbox{Composite-B16} observations use data from two sources, namely NSO/SP and SOLIS/ISS. For this cycle, the residual between \mbox{Composite-B16} and NSO/SP (blue) are also higher.

To characterise the behaviour of the \sind{} on the solar activity cycle timescale, we consider the quantities such as values at cycle minimum ($S_{\rm min}$), maximum ($S_{\rm max}$), cycle-average ($\langle S\rangle$) and the amplitude ($\Delta S$). While there are different ways to define them (e.g. to fit cycle shape functions such as skewed Gaussians \citep[][see also \citealt{2011SoPh..273..231D}]{2017ApJ...835...25E} or functions of the type used in \citealt{1994SoPh..151..177H}), we follow the approach by \citet{2017ApJ...835...25E}  which allows a direct comparison of our estimates with theirs and use a skewed Gaussian cycle shape model function defined as: 
\begin{equation}
    S_{\rm fit}(t) = \Delta S\ {\rm exp}\ \bigg(-\frac{(t-t_{\rm m})^2}{2b^2[1+a(t-t_{\rm m})^2]}\bigg)\ +\ S_{\rm min}\ ,
    \label{eq:cycleshape}
\end{equation}
where $t$ is the time grid covering a given cycle, $\Delta S$ is the amplitude of the cycle, $t_{\rm m}$ is the time of the maximum of the cycle, $b$ is the width of the rising phase of the cycle, and $a$ is the cycle asymmetry parameter.

We fit the function given by \eqn{eq:cycleshape} to both the \mbox{Composite-B16} (red curves in \fref{fig:sindexbersatnew}a) and to our model calculations (purple curves in \fref{fig:sindexbersatnew}a). We do not include the solar cycle 22 into the fit because of the significant gap in the magnetograms used in our model. The values of the cycle parameters determined for \mbox{Composite-B16} and our model are listed in \tab{tab:sparams}. Note that $S_{\rm min}$ is the minimum (in this case corresponding to the minimum at the end of the cycle), $S_{\rm max}$ is the maximum and $\langle S \rangle$ is the mean of the cycle shape fit. We remark that these parameter values depend on the start time of a cycle. We used the cycle start times from \citet{2017ApJ...835...25E} in order to make a direct comparison with their estimates (see \tab{tab:sparams}). The \mbox{Composite-B16} and our model display very similar minimum and maximum values for both cycles considered. The averages for cycle 21 are slightly different due to a different time sampling of the \mbox{Composite-B16} and our model. The cycle parameters found for both \mbox{Composite-B16} and our model are in close agreement with the \mbox{Composite-E17} estimates for the corresponding solar cycles (see \tab{tab:sparams}).

\begin{figure*}[ht!]
    \centering
    \includegraphics[scale=0.9,trim=0.5cm 6.0cm 0.cm 0.cm,clip]{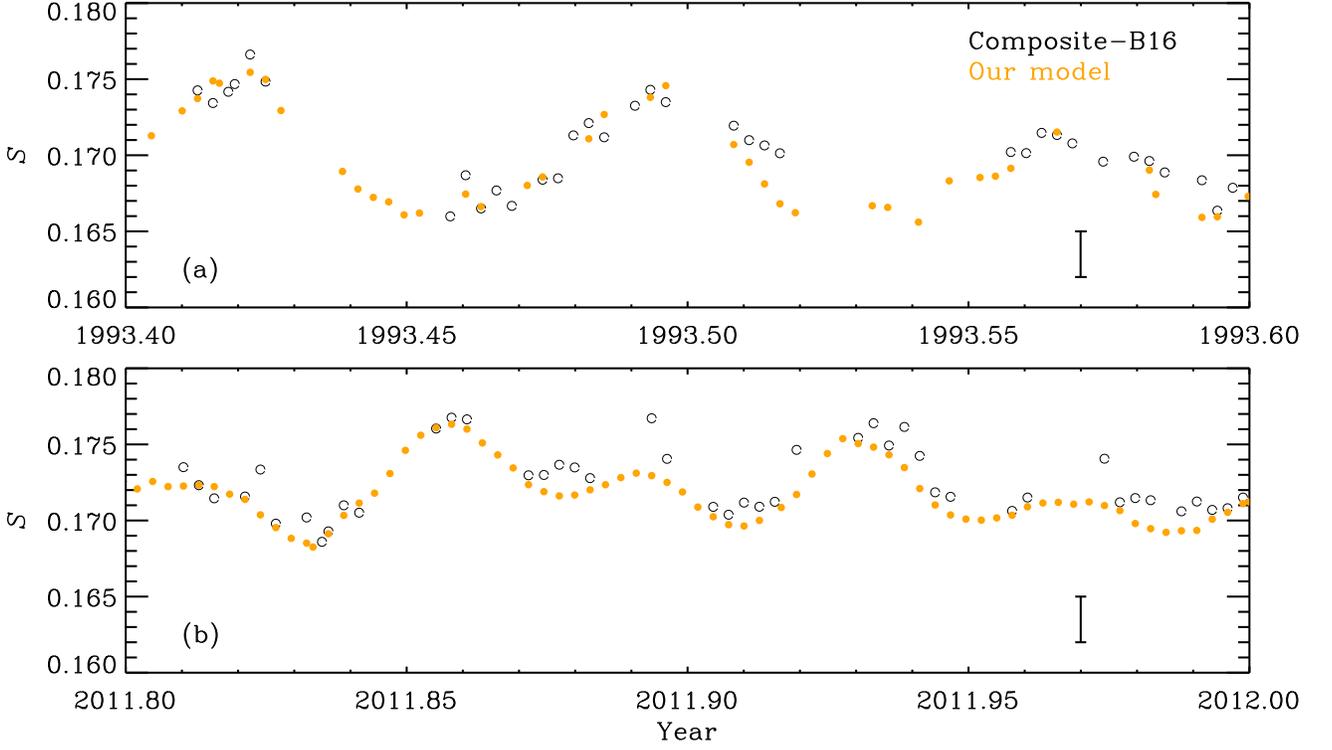}
    \caption{Comparison between the measured \sind{} from the \mbox{Composite-B16} (black circles) and the \sind{} reconstructed with our model (orange circles) for a few rotational cycles. The error bars shown in each panel represent the $3\sigma$-uncertainty, where $\sigma$ is the standard deviation of the \mbox{Composite-B16} values in the period $2008.5-2009.5$.}
    \label{fig:sindexbersatrot}
\end{figure*}

\begin{figure}
    \centering
    \includegraphics[scale=0.5]{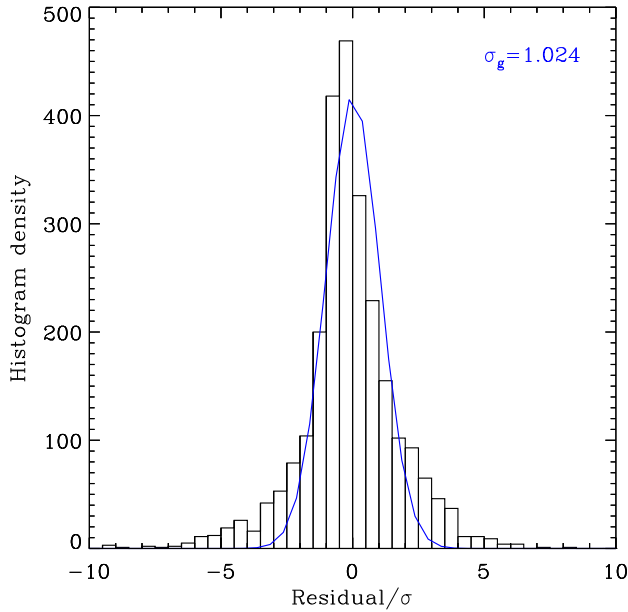}
    \caption{Histogram of the residual (defined as the difference between the same-day \sind{} values from Composite-B16 and our model) normalized to the standard deviation ($\sigma$) of the \mbox{Composite-B16} values in the period $2008.5-2009.5$ introduced in \fref{fig:sindexbersatrot}. The blue curve is the Gaussian fit to the distribution with $\sigma_{\rm g}$ being the standard deviation of the fitted Gaussian function.}
    \label{fig:residual}
\end{figure}

\begin{table*}[ht!]
\centering
\caption{Parameters of the $S$-index activity cycles 21 and 23. }
\begin{tabular}{c|ccc|ccc}
\midrule[1.5pt]
\multirow{2}{*}{Parameters} & \multicolumn{3}{c}{Cycle 21} &  \multicolumn{3}{c}{Cycle 23} \\
& \mbox{Composite-E17} & \mbox{Composite-B16} & Our model  & \mbox{Composite-E17} & \mbox{Composite-B16} & Our model \\ 
\midrule[1.5pt]
$S_{\rm min}$ & 0.1629 & 0.1636 & 0.1636 & 0.1634 & 0.1623 & 0.1621\\
$S_{\rm max}$ & 0.1790 & 0.1789 & 0.1792 & 0.1780 & 0.1773 & 0.1778\\
$\langle S\rangle$ & 0.1717 & 0.1692 & 0.1723 & 0.1701 & 0.1691 & 0.1700\\
$\Delta S$ & 0.0164 & 0.0152 & 0.0157 & 0.0143 & 0.0149 & 0.0156\\
\bottomrule[1.5pt]
\end{tabular}
\label{tab:sparams}
\end{table*}

\begin{table*}[ht!]
\centering
\caption{log($R^\prime_{\rm HK}$) values for solar cycles 21 and 23.}
\begin{tabular}{c|ccc|ccc}
\midrule[1.5pt]
\multirow{2}{*}{Parameters} & \multicolumn{3}{c}{Cycle 21} &  \multicolumn{3}{c}{Cycle 23} \\
& \mbox{Composite-E17} & \mbox{Composite-B16} & Our model & \mbox{Composite-E17} & \mbox{Composite-B16} & Our model \\ 
\midrule[1.5pt]
log($R^\prime_{\rm HK,min}$) & -4.979 & -4.975 & -4.975 & -4.976 & -4.982 & -4.984\\
log($R^\prime_{\rm HK,max}$) & -4.891 & -4.893 & -4.892 & -4.899 & -4.901 & -4.898\\
log($\langle R^\prime_{\rm HK}\rangle$) & -4.930 & -4.943 & -4.926 & -4.938 & -4.944 & -4.939\\
log($\Delta R^\prime_{\rm HK}$) & -5.626 & -5.658 & -5.648 & -5.686 & -5.666& -5.646\\
\bottomrule[1.5pt]
\end{tabular}
\label{tab:rhkparams}
\end{table*}

The \sind{} is not an ideal proxy for comparing magnetic activity of different stars, due to its dependence on the effective temperature. Both the flux in the R and V passbands, and the \ca{} emission are affected by the stellar effective temperature \citep{1970PASP...82..169L,1984ApJ...279..763N}. For a more accurate comparison of the magnetic activity of stars with different effective temperatures, the photospheric contribution to the chromospheric emission must be removed and the dependence of the R and V fluxes on the effective temperature must be taken into account. For this reason, instead of the \sind{}, many studies involving stars with different effective temperatures use the derivative index $R^\prime_{\rm HK}$ \citep[][see also \sref{ssec:anomaly}]{1984ApJ...279..763N}, defined as:
\begin{equation}
    R^\prime_{\rm HK} = R_{\rm HK}\ -\ R_{\rm phot}\ ,
    \label{eq:rhkp}
\end{equation}
where $R_{\rm HK}=1.34\times 10^{-4}\ C_{cf} \ S$ and $R_{\rm phot}$ is the photospheric correction. Here, $C_{cf}$ is the color ($B-V$) dependent factor  used to convert \sind{} to $R_{\rm HK}$. It corrects for the temperature dependence of the R and V fluxes and is defined as:
\begin{equation}
\begin{split}
    {\rm log} C_{cf}& = 1.13\ (B-V)^3 - 3.91\ (B-V)^2 \\ & + 2.84\ (B-V) - 0.47\ .
\end{split}
    \label{eq:logcf}
\end{equation}
$R_{\rm phot}$ is calculated from the $(B-V)$ color using
\begin{equation}
\begin{split}
    {\rm log} R_{\rm phot} & =  -2.893\ (B-V)^3 \\ & + 1.918\ (B-V)^2 - 4.898 \ .
\end{split}
    \label{eq:logrphot}
\end{equation}
The values of log($R^\prime_{\rm HK}$) computed using Eqs.~(\ref{eq:rhkp})-- (\ref{eq:logrphot}) and $(B-V)=0.653$ following \citet{2017ApJ...835...25E} are shown in \tab{tab:rhkparams}. We use the procedure outlined here to derive the chromospheric emission variations of the Sun and compare the Sun with other lower main-sequence stars in \sref{ssec:anomaly}.

A comparison between the modeled daily \sind{} values and the \mbox{Composite-B16} is shown for a few rotational cycles in \fref{fig:sindexbersatrot}. Since \citet{2016SoPh..291.2967B} do not provide any measurement uncertainties we choose the scatter ($\sigma$) in the \mbox{Composite-B16} values between $2008.5-2009.5$ (when our model returns very low \sind{} variability) as a representative of the uncertainty and show the $3\sigma$ (with $\sigma = 0.001$) value as error bar in \fref{fig:sindexbersatrot}. One can see that our model values agree well with those from the \mbox{Composite-B16} also on the rotational timescale. In \fref{fig:residual} we show the histogram of the residual (the difference in the same-day \sind{} values from \mbox{Composite-B16} and our model up to 2012) normalized to the estimated uncertainty ($\sigma$ introduced earlier) in \mbox{Composite-B16}. The blue curve shows the fit to the histogram obtained assuming a normal distribution with a standard deviation $\sigma_{\rm g}$. Although not ideal, the Gaussian fit to the distribution looks reasonable and has a width very close to unity as indicated ($\sigma_{\rm g}=1.024$). The distribution in black has stronger wings, suggesting that there are more outliers than expected from a strictly normal distribution. Instead of the 99.7\%\ expected from a pure Gaussian, 90\%\ of the residuals plotted in \fref{fig:residual} lie within $\pm 3\sigma_{\rm g}$. The relatively high number of outliers could be a result of the inaccuracies in our model (e.g., assumption of a constant expansion factor $A_{\rm exp}$, neglecting the contribution from spots, and a slight mismatch between the observed and modeled spectra) and/or problems in the \sind{} data.

\begin{figure*}[ht!]
    \centering
    \includegraphics[scale=0.9]{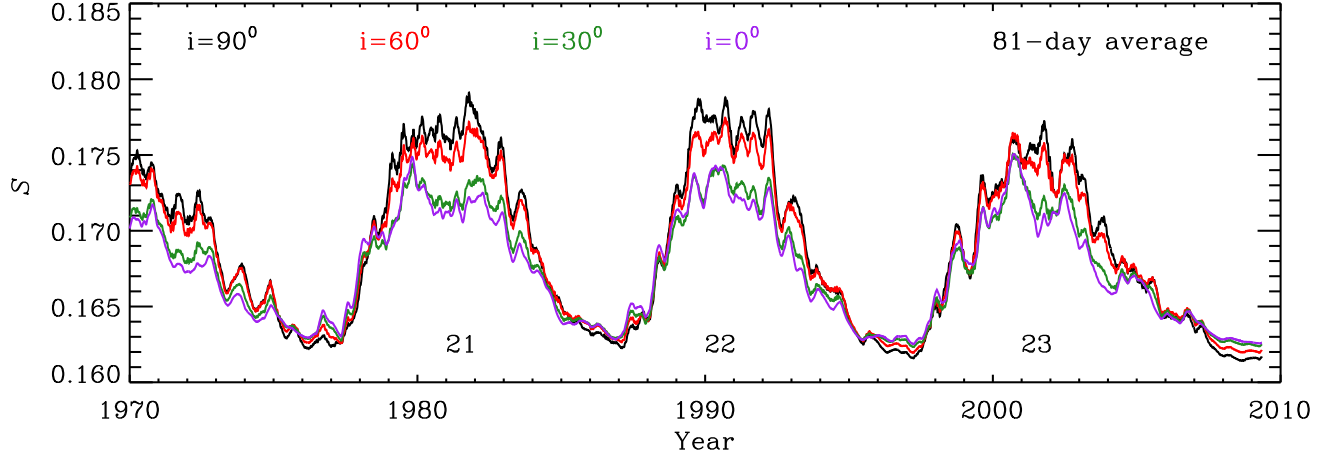}
    \caption{81-day running averages of the \sind{} during solar cycles $21-23$ for i = 90\textdegree\ (equator-on view; black), i = 60\textdegree\ (red), i = 30\textdegree\ (green), and i = 0\textdegree\ (pole-on view; purple).}
    \label{fig:13mnavg}
\end{figure*}
\begin{figure*}[h!]
    \centering
    \includegraphics[scale=0.48,trim=0.3cm 0.cm 1.5cm 0.cm,clip]{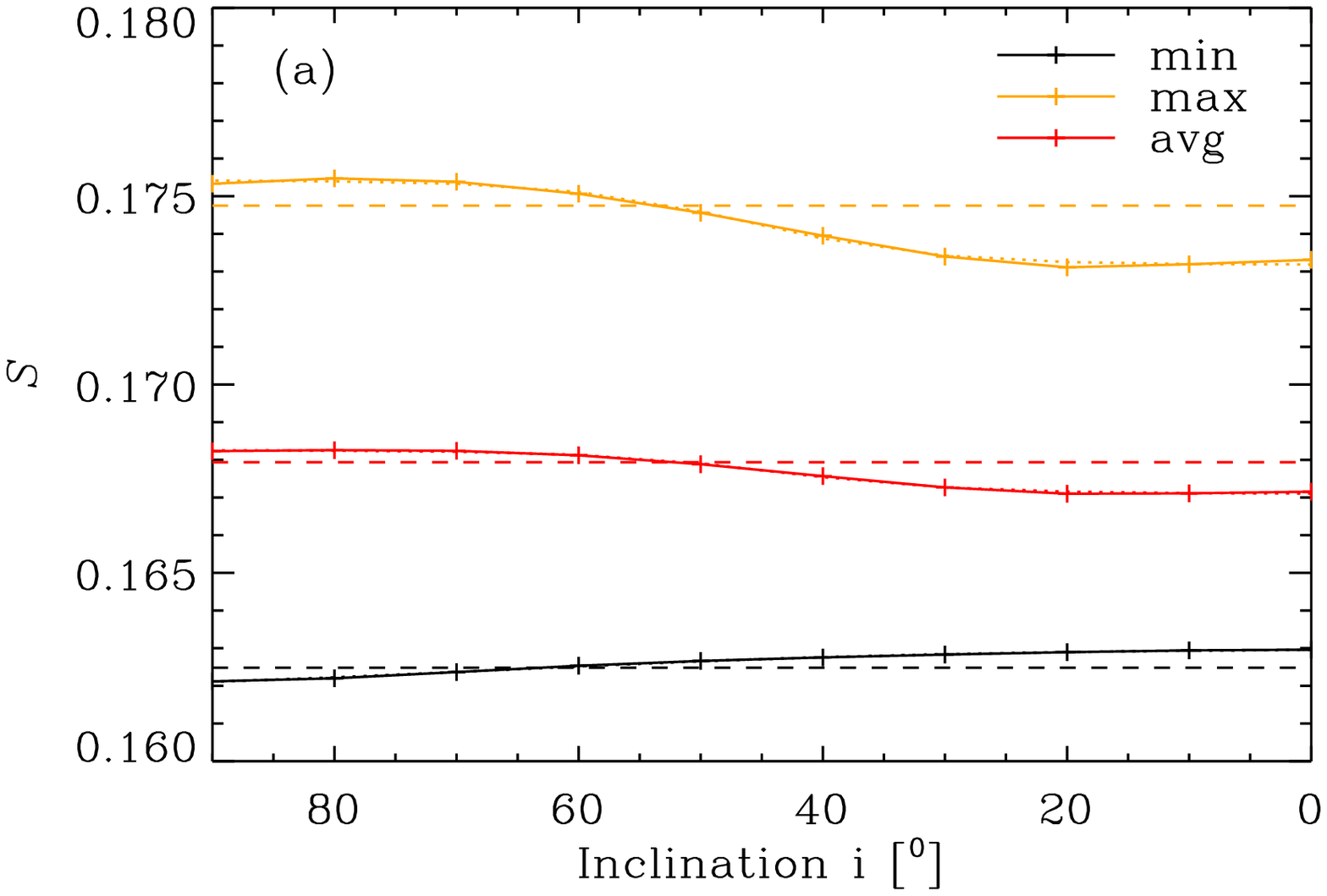}
    \includegraphics[scale=0.48,trim=0.3cm 0.cm 1.5cm 0.cm,clip]{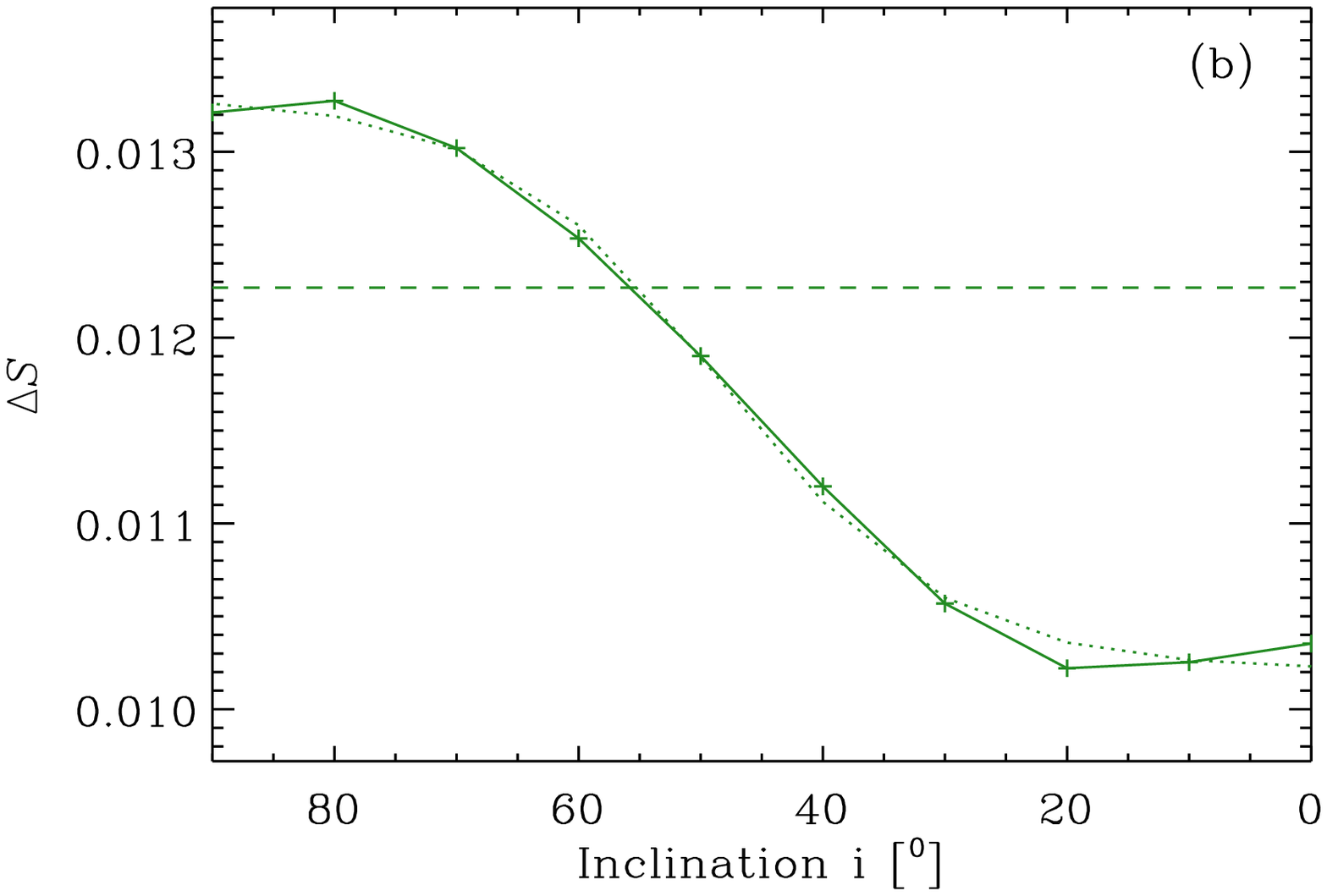}
    \caption{Dependence of the \sind{} on the inclination for solar cycle 23. Panel (a): \sind{} at the cycle minimum (black symbols), cycle maximum (orange symbols) and cycle average (red symbols). Panel (b): cycle amplitude. The dashed lines indicate the inclination averaged values. The dotted lines show the fits to the inclination dependent curves obtained using the function in \eqn{eq:samp}.}
    \label{fig:cyclets}
\end{figure*}

\section{Effect of the inclination}
\label{sec:incl}
So far we focused on the Sun as observed from the ecliptic plane, corresponding to an inclination of 90\textdegree\ (the 7.25\textdegree\ angle between the solar equator and the ecliptic has been neglected, which leads to a very weak periodic half-yearly difference between the observed and modeled \sind{}). In this section we explore the impact of the inclination on the solar \sind{}. Since solar active regions tend to emerge within the so-called activity belts (e.g., most of the spots emerge at latitudes between $\pm5\degree$ and $\pm30\degree$), a change in the inclination angle strongly affects the distribution of active regions on the visible solar disk. To quantify such changes, we have to resort to numerical simulations, as out-of-ecliptic observations are not available for the Sun.

To this end, we follow the approach described in \citet{Nina2, Nina1}, who used a surface flux transport model \cite[SFTM; in the form introduced by][]{2010ApJ...719..264C}. The SFTM is an advective-diffusive model, which describes the passive transport of the radial component of the magnetic flux on the stellar surface under the influence of the differential rotation and the meridional flow. In the SFTM, the magnetic flux emerges in the form of bipolar active regions. The emergence characteristics of these regions are determined using the semi-empirical sunspot-group record of \citet{2011A&A...528A..82J} for the period $1700-2010$. The statistical properties of sunspots (e.g., tilt-angles, latitudinal distribution etc.) in this semi-empirical sunspot-group record reflect those of the sunspot record of the Royal Greenwich Observatory. In \citet{Nina1}, the latitudes, numbers, and tilt-angle distributions of the emergences from \citet{2011A&A...528A..82J} are preserved, but the active regions are made to emerge at random longitudes. This randomization is essential to circumvent asymmetries between the near- and far- side of the Sun (for an Earth-bound observer) and thus is ideal for studying the inclination effect, on both the rotational and the activity cycle timescales \citep[for further details, see][]{Nina2,Nina1}.

In this section we use the same approach as in \sref{sec:svar} to compute the \sind{}, but replace the fractional area coverages in \fref{fig:ffactors} deduced from ecliptic-bound observations with those calculated by \citet{Nina1} for the period $1700-2010$, which can be re-sampled to various inclinations. As described in \sref{ssec:rh} we neglect the sunspot contribution to the \sind{}, so that the effect of inclination is fully attributed to the changes in fractional area coverage of faculae on the visible solar disk and the CLV of the facular contrast in H, K, R, and V passbands (see \fref{fig:contrast}). Throughout the rest of the paper, we denote the inclination angle as `i' with i = 90\textdegree\ representing the equator-on view and i = 0\textdegree\ the North pole-on view.

While comparing the fractional disk area coverages of faculae obtained from the SFTM for i = 90\textdegree\ with those shown in \fref{fig:ffactors}, we noticed that the SFTM values were on an average 2\%\ lower relative to the observed fractional areas in \fref{fig:ffactors}. This is because of the absence of the ephemeral regions in the SFTM model we used \citep[see a detailed discussion in][]{Dasi2016}. An offset between the observed \sind{} values and those computed from our model was introduced by the missing ephemeral regions, so that the \sind{} from our model were slightly lower than the observed one. The amplitudes of the cycles in the \sind{}, however, remained comparable to the observed ones. To compensate for the reduced \sind{} values due to the missing ephemeral regions, we scaled the calibration constant $\alpha_{\rm c}$ until we closely matched the observed \sind{} values shown in \fref{fig:sindexbersatnew}. The new value of the calibration constant is $\alpha_{\rm c}=2.53$, which holds for all the results presented in this section.

\subsection{Variations on the activity cycle timescale}
\label{ssec:svaract}
In \fref{fig:13mnavg}, we show the 81-day smoothed means of the \sind{} values obtained for the solar cycles $21-23$ for different inclinations. The amplitude of the \sind{} variation over the activity cycle decreases with decreasing inclination (i.e. when going from the ecliptic to pole-on view). There are two effects which contribute to this decrease. First, the fraction of the visible disk covered by faculae decreases when the Sun is viewed from out of the ecliptic plane. Indeed, solar active regions preferably emerge at low and intermediate latitudes. Consequently, a shift of the observer from the equatorial plane leads to active regions emerging closer to the visible limb, so that their visible areas are reduced by the foreshortening effect. Second, as shown in \fref{fig:contrast}, while the facular contrast in the line core passbands K and H increases strongly towards the limb, the magnitude of the increase is relatively small (e.g., in comparison to the increase of the contrast in pseudo-continuum bands R and V). Such a moderate increase of the contrast cannot compensate the foreshortening effect so that the shift of facular regions towards the limb reduces the emission in the K and H passbands, while having only a minor effect on R and V passbands. This behaviour leads to a decrease of the facular contribution to \sind{} (which is a disk integrated quantity) and, consequently, the amplitude of the \sind{} variability on the activity cycle timescale. We note an opposite behaviour of photometric variability in the visible spectral domain (e.g. measured in the Str{\"o}mgren filters) with inclination \citep[see, e.g.][]{2001A&A...376.1080K, 2014A&A...569A..38S, Veronika_2018}. The facular contrast in the visible spectral domain  strongly increases towards the limb (similarly to the increase of the contrast in pseudo-continuum bands R and V, see left panel of \fref{fig:contrast}), overcompensating the foreshortening effect. As a result, the amplitude of the activity cycle in Str{\"o}mgren filters increases with decreasing inclination. Interestingly, \fref{fig:13mnavg} shows that during the activity minimum, the \sind{} increases slightly when the star is observed from closer to the poles. This is attributed to the increased contribution from the polar network to the \sind{}. The poleward migration of the magnetic fields to latitudes above $\pm60\degree$ leads to the build up of the polar flux \citep{2015LRSP...12....4H}.

Focusing on cycle 23, we now quantify the \sind{} dependence on the inclination in more detail. For this cycle, we define the minimum and maximum values of the \sind{} as annual averages over the year 1996 and over the period $2000.5-2001.5$, respectively. The cycle 23 minimum and maximum values for various inclinations are shown as black and orange symbols, respectively, in \fref{fig:cyclets}a. The cycle averaged values, defined as the average over daily values during the period 1996--2009 for a given inclination, are shown by red symbols. In \fref{fig:cyclets}b, we show the amplitude of the \sind{} variations, defined as the difference between the cycle maximum and cycle minimum \sind{} values. The maximum \sind{} value over the activity cycle decreases gradually as the inclination changes from the equator-on case to the pole-on case. This decrease is attributed to the decrease in the coverage of the disk visible to the observer by faculae. However, the trend is opposite for the minimum (also visible in \fref{fig:13mnavg}). The \sind{} minimum values increase very slowly due to the enhanced amount of kG field magnetic flux in the polar regions. Such polar magnetic field concentrations forming polar network contribute to the area coverage by plage, thus leading to an increase in the \sind{}. The cycle averages follow the same trend as the cycle maximum, as does the amplitude of the \sind{} variations. The horizontal dashed lines show the values averaged over all inclinations computed using \citep[as done by][]{Nina2}
\begin{equation}
    S_{\rm incl\_avg}\ =\ \frac{\sum_{\rm i=0}^{90} S_{\rm i}\ \rm{sin(i)}}{\sum_{\rm i=0}^{90} \rm{sin(i)}}\ ,
    \label{eq:wtsigs}
\end{equation}
where the weighting factor \rm{sin(i)} corresponds to the probability that a star is observed at inclination `i'. Interestingly, the intersection of the inclination-dependent curves with those averaged over inclinations occurs very close to ${\rm i}=57\degree$, in spite of the trend in the \sind{} variation being opposite in the cycle maximum and minimum. This value of $57\degree$ is often used in the literature to characterize the intermediate case between equator-on and pole-on views \citep{1993JGR....9818907S,1998ApJS..118..239R,2001A&A...376.1080K}. All in all, \fref{fig:cyclets} shows that the inclination averaged \sind{} is well represented by the \sind{} at ${\rm i}=57\degree$. The simplification of using \sind{} at ${\rm i}=57\degree$ therefore appears to be reasonable when dealing with a sample of stars having a random distribution of inclinations.

The dependence of the cyclic variability of the \sind{} as well as the absolute value of the \sind{} on inclination can be approximated by a logistic function of the form
\begin{equation}
    Q({\rm i})\ =\ {\rm A} + \frac{{\rm B}}{1\ +\ {\rm exp}\ [{-{\rm C}({\rm i}\ +\ {\rm D})]}}\ ,
    \label{eq:samp}
\end{equation}
where A, B, C, D are constants and $Q({\rm i})=\{S_{\rm min}({\rm i}),S_{\rm max}({\rm i}),S_{\rm avg}({\rm i}),\Delta S({\rm i})\}$. The dotted lines in \fref{fig:cyclets} show the fits to the inclination dependent curves obtained with this function. The best fit values of the constants in \eqn{eq:samp} for the four curves in \fref{fig:cyclets} are given in \tab{tab:fitparam}.

\begin{table*}[ht!]
\centering
\caption{Best fit values of the constants in \eqn{eq:samp} corresponding to the dotted lines in \fref{fig:cyclets}.}
\begin{tabular}{c|cccc}
\midrule[1.5pt]
Parameter & A & B & C & D\\
\midrule[1.5pt]
S$_{\rm min}$ & $+0.162$ & $-0.001$ & $+0.052$ & $-66.617$\\
S$_{\rm max}$ & $+0.173$ & $+0.002$ & $+0.132$ & $-46.025$\\
S$_{\rm avg}$ & $+0.167$ & $+0.001$ & $+0.130$ & $-43.703$\\
$\Delta S$ & $+0.010$ & $+0.003$ & $+0.106$ & $-48.288$\\
\bottomrule[1.5pt]
\end{tabular}
\label{tab:fitparam}
\end{table*}

\subsection{Variations on the rotational timescale}
\label{ssec:svarrot}
\begin{sidewaysfigure}[ht!]
    \centering
    \includegraphics[angle=270,width=\textwidth,totalheight=0.6\textheight,trim=2.5cm 0.cm 3.0cm 0.cm,clip]{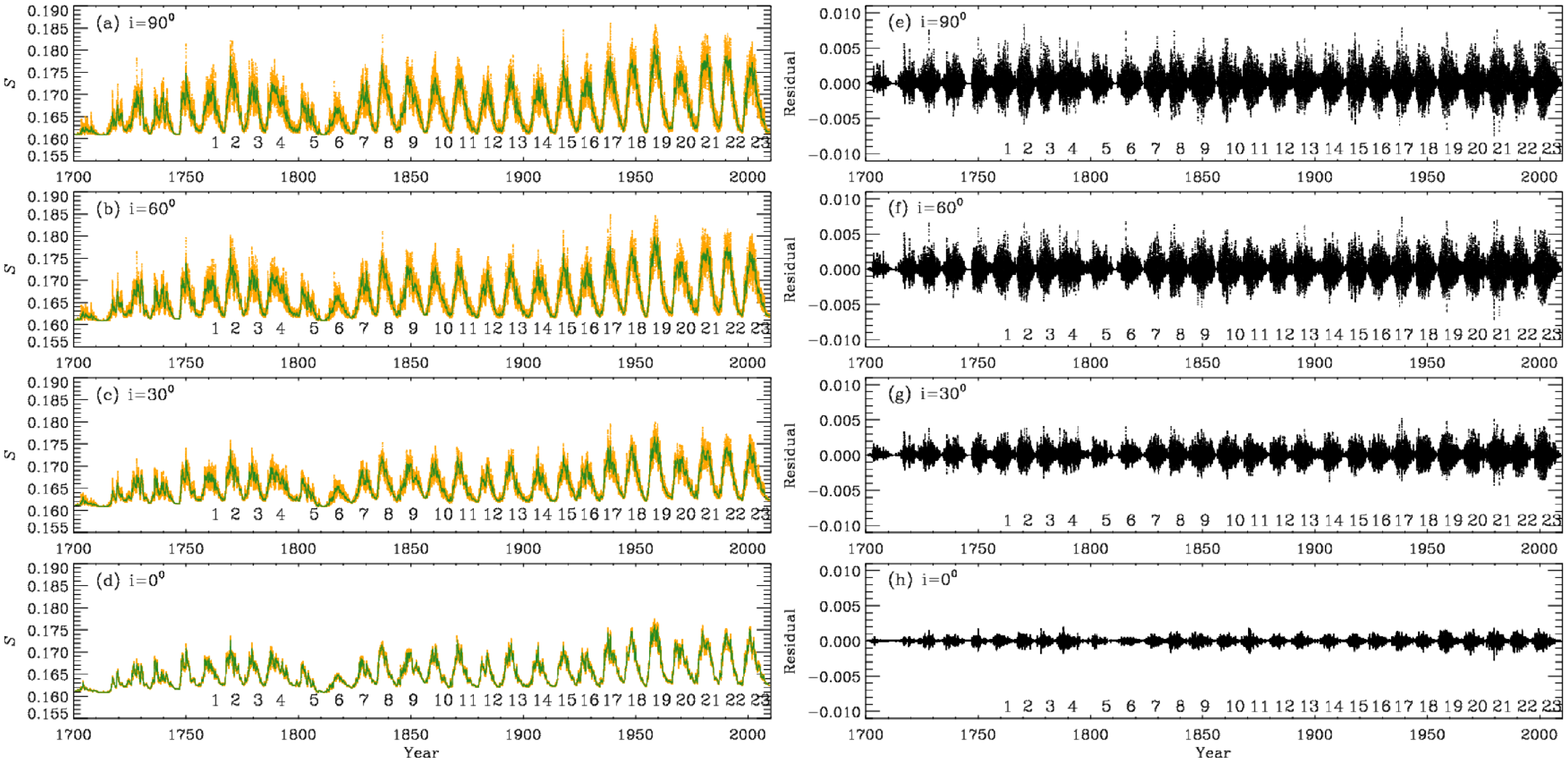}
    \caption{Simulated time series of the \sind{} back to 1700 for inclination i = 90\textdegree. Panels (a-d): daily values of the \sind{} (orange dots) and the 81 day running mean values (green curve). Panel (e-h): the difference between the daily and 81 day smoothed \sind{} values (residual). Solar cycle numbers are indicated at the bottom of each panel.}
    \label{fig:sind90}
\end{sidewaysfigure}

In \fref{fig:sind90}, we show the time series of the \sind{} going back to 1700 for inclinations of 90\textdegree, 60\textdegree, 30\textdegree, and 0\textdegree. These figures have the same y-scale to facilitate a visualization of the inclination effects. The orange dots in the upper panels are the daily \sind{} values and the green curves are the corresponding 81-day running averages. The residuals i.e. the difference between the daily and the 81-day smoothed values ($S_{\rm daily}-S_{\rm 81d-mean}$), representing the \sind{} variability on the rotational timescale, are plotted in the lower panels. The numbers at the bottom of each panel mark the number of the solar cycle. The Figure suggests that stronger cycles (such as cycle 19), during which the number of active regions emerging at the solar surface is higher, show stronger \sind{} variations compared to weaker cycles (such as cycle 6), which witness the emergence of relatively less active regions. A few outliers seen in the residuals may correspond to large active region.

It is evident that, while the amplitudes of both the activity cycle and the rotational variability in the \sind{} decrease as the inclination decreases, the drop in the amplitude of the rotational variability is much stronger. In particular, for the pole-on view (i = 0\textdegree), the scatter in the residual is much lower. This is because the rotation induced changes in the fractional coverage by faculae decreases as the inclination decreases and the variation produced by rotation disappears at ${\rm i}=0^\degree$ \citep[see e.g. Figure 7 of][]{Nina2}. At ${\rm i}=0^\degree$ the faculae simply move along in circles staying at roughly the same limb distances instead of crossing all $\mu$ values as at ${\rm i}=90^\degree$. In order to better illustrate this decline in the \sind{} rotational variability, we use the metric $S_{30}$, similar to the R$_{30}$ metric used to quantify the photometric variability \citep{2013ApJ...769...37B}. To compute $S_{30}$, we divide the time series of daily \sind{} values over the period $1755-2010$ into 30-day segments. We then take the maximum and minimum \sind{} values within each segment and divide their difference by the mean value in the segment. Since our calculations do not suffer from the measurement noise (unlike observations), we directly take the difference between the extrema, instead of taking the difference between the 5th and 95th percentile values in the segment as done by \citet{2013ApJ...769...37B}.

\begin{figure*}[ht!]
    \centering
    \includegraphics[scale=0.52,trim=1.2cm 0.5cm 1.8cm 0.cm,clip]{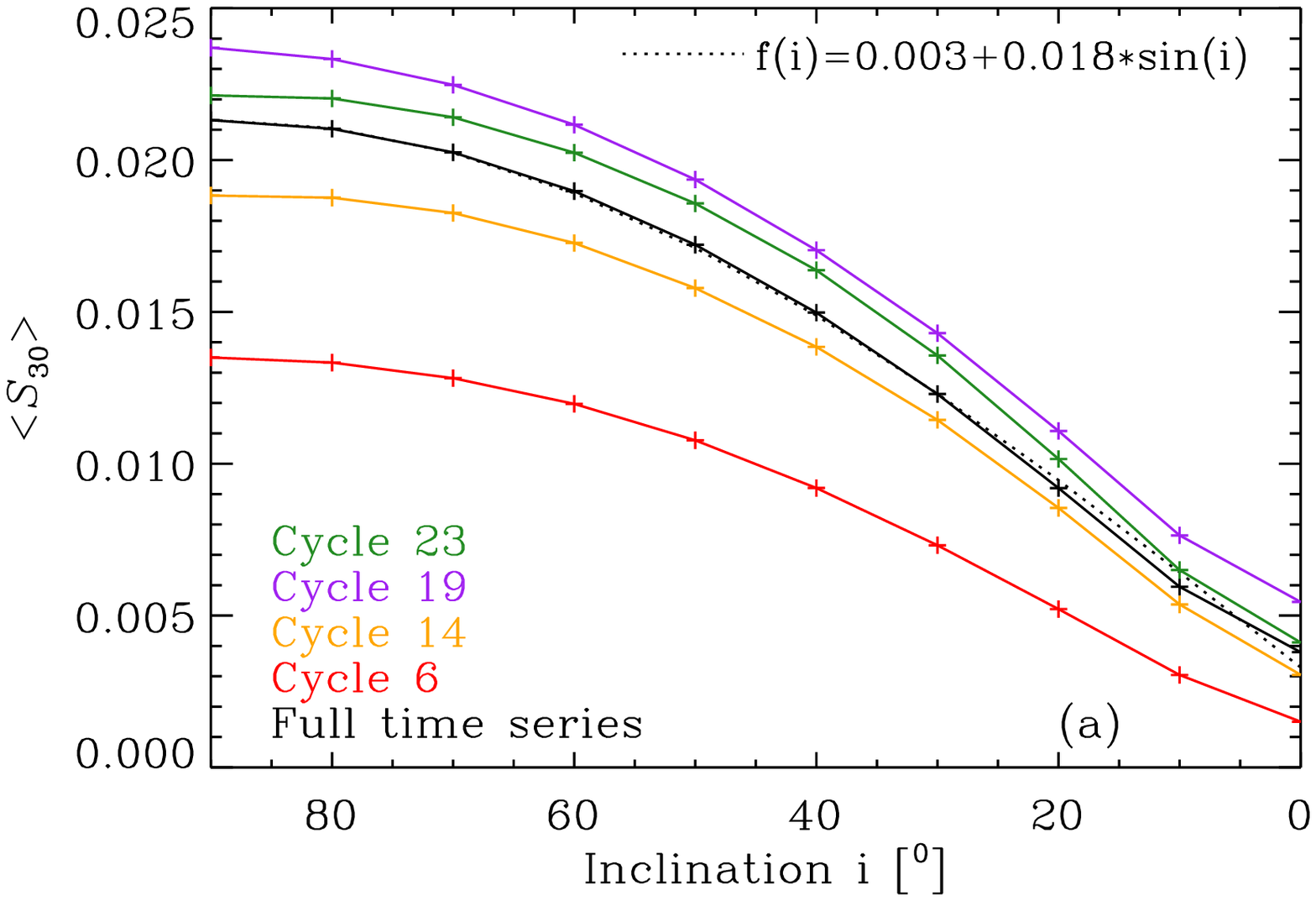}
    \includegraphics[scale=0.52,trim=1.2cm 0.5cm 1.8cm 0.cm,clip]{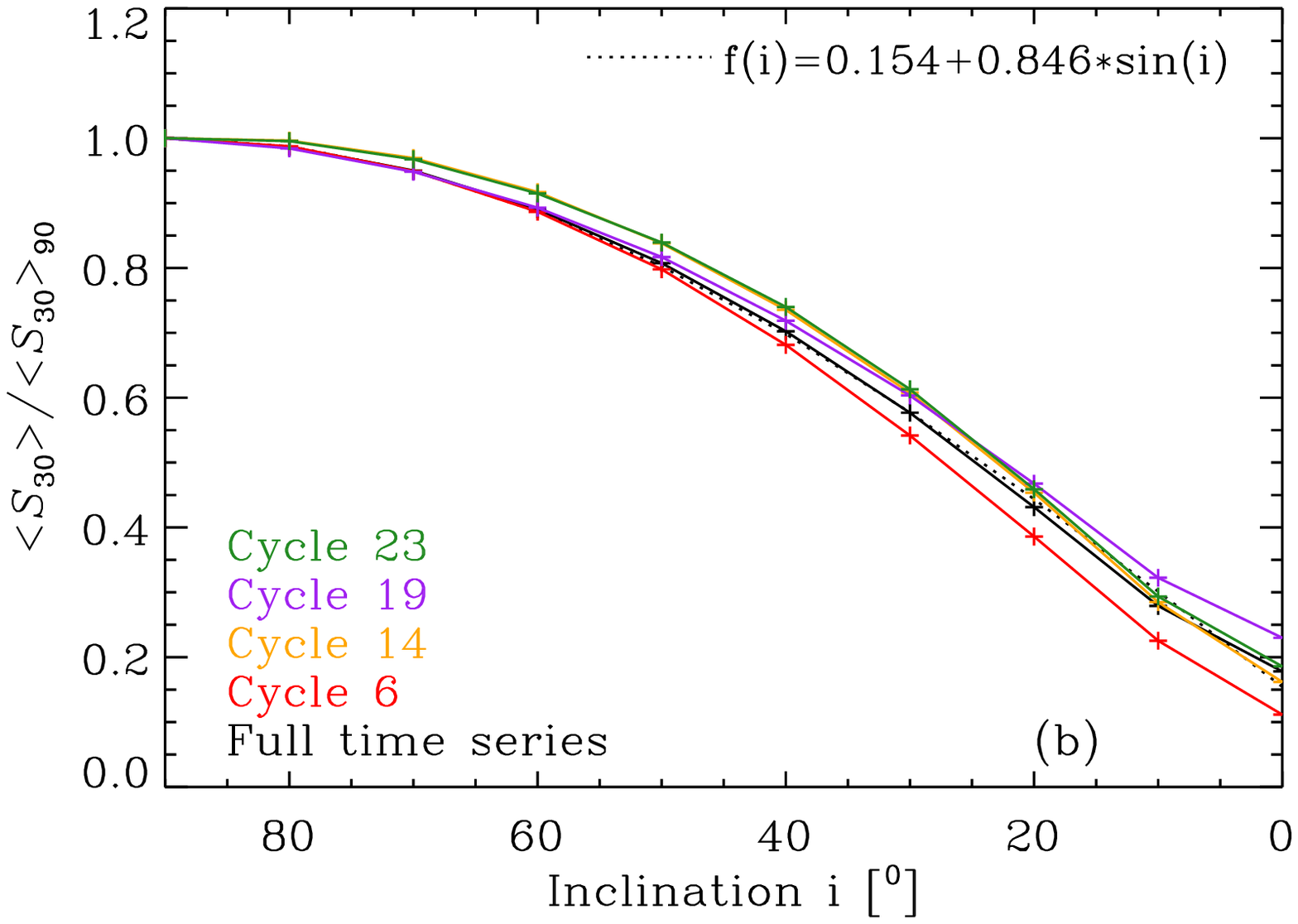}
    \caption{\sind{} variability on the rotational timescale, expressed in terms of $\langle S_{30}\rangle$ (panel a) and the variability normalized to the respective value at ${\rm i}=90\degree$ (panel b), averaged over all 23 cycles (black), cycle 6 (red), cycle 14 (orange), cycle 19 (purple), and cycle 23 (green). Dotted black curves are the fits to the full time series in black obtained using the functions indicated in the panels.}
    \label{fig:rotats}
\end{figure*}

\begin{figure*}[hb!]
    \centering
    \includegraphics[scale=0.5,trim=1.2cm 0.5cm 1.0cm 0.1cm,clip]{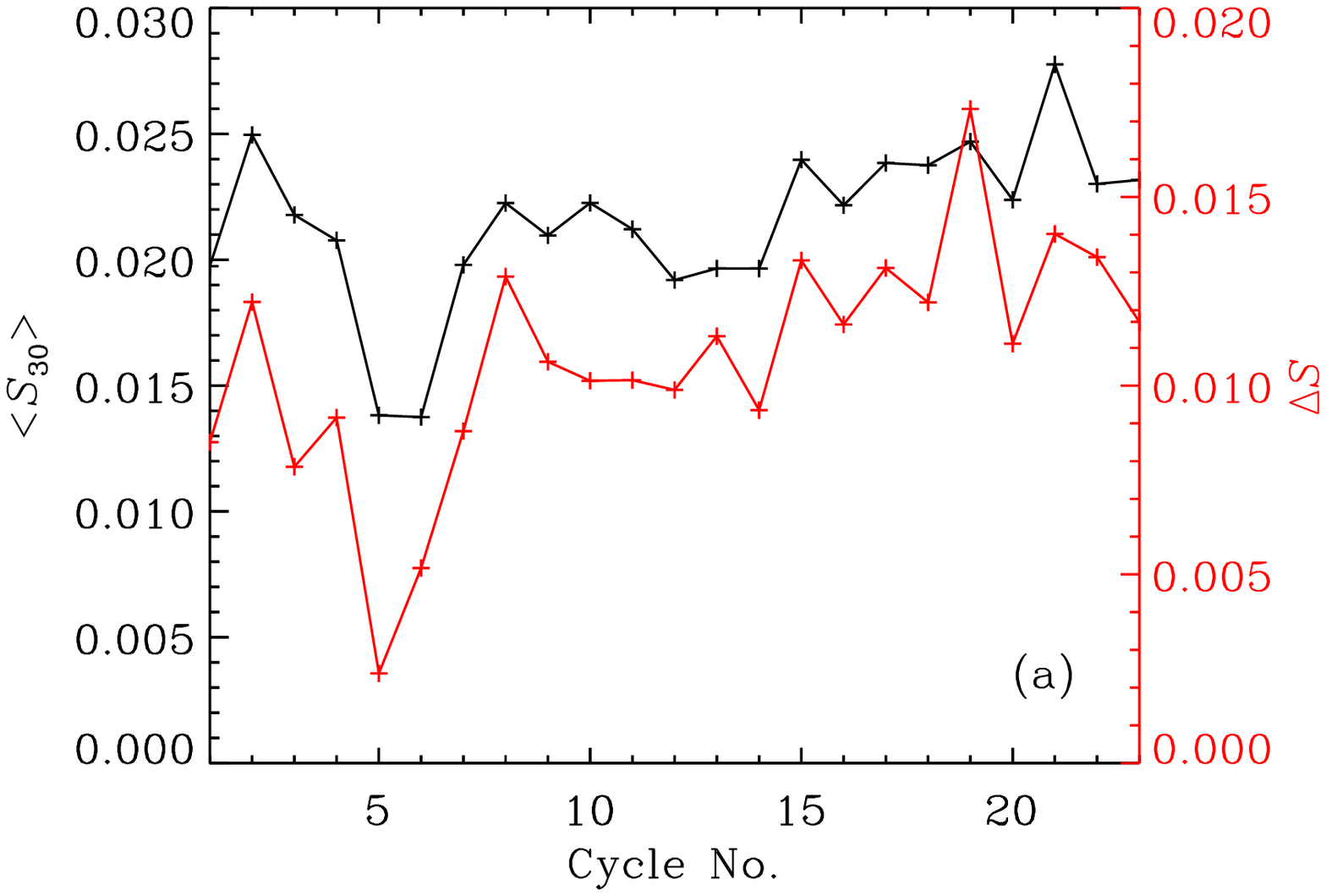}
    \includegraphics[scale=0.5,trim=1.2cm 0.5cm 3.0cm 0.1cm,clip]{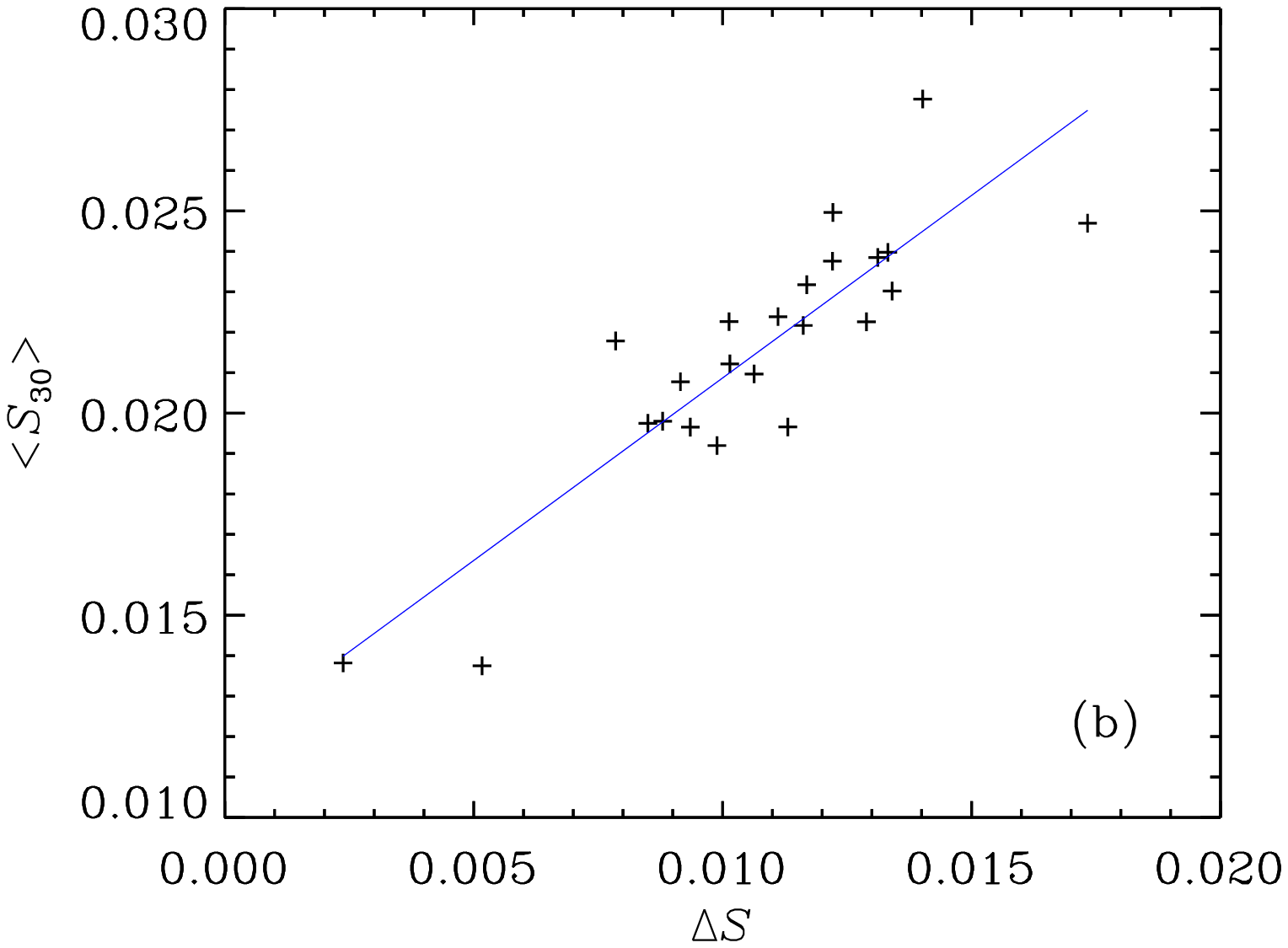}
    \caption{Panel (a): variations of $\langle S_{30}\rangle$ (black) and $\Delta S$ (red) across 23 solar cycles and for ${\rm i}=90\degree$. Panel (b): scatter plot of $\langle S_{30}\rangle$ and $\Delta S$ values shown in panel (a), with the blue line showing the linear fit to the data. The linear Pearson correlation coefficient is 0.87.}
    \label{fig:cycle}
\end{figure*}

In \fref{fig:rotats}a, we plot the time averaged rotational variability denoted by $\langle S_{30}\rangle$ as a function of inclination for cycle 6 (i.e. $1810-1822$, red), cycle 14 ($1902-1913$, orange), cycle 19 ($1954-1964$, purple), and cycle 23 ($1996-2009$, green). The black curve labeled as full time series is obtained by computing $\langle S_{30}\rangle$ over the period $1755-2010$, covering cycles $1-23$ (see \fref{fig:cycle} for the variation of $\langle S_{30}\rangle$ corresponding to ${\rm i}=90\degree$ across all 23 solar cycles). We neglected the values before 1755 as the sunspot record in that period is less reliable due to the scarcity of sunspot number data \citep[see e.g.][for more details]{2017A&A...602A..69C,2019NatAs...3..205M}. \fref{fig:rotats}a shows that the mean level of the rotational variability decreases more strongly (by about 81\%\ for cycle 23) with decreasing inclination than the reduction in the variability over the solar cycle (which is about 22\%, see \fref{fig:cyclets}). In addition, we see that $\langle S_{30}\rangle$ does not vanish for ${\rm i}=0\degree$ in spite of solar rotation not contributing to the variability when the solar disk is viewed pole-on. This is due to the emergence and the evolution of faculae on the visible solar disk during their lifetime, which is typically days to months. Stronger cycles show a higher level of the variability than weaker cycles, and the level of the variability for the full $1700-2010$ time series lies between that of the two intermediate strength cycles 14 and 23. Figure~\ref{fig:rotats}b shows the $\langle S_{30}\rangle$ values normalized to the corresponding values at ${\rm i}=90\degree$. The differences in the inclination effect between the cycles (e.g. compare red and purple curves in \fref{fig:rotats}b) are due to the fact that stronger cycles reach to higher latitudes \citep[e.g.][]{2008A&A...483..623S} and the faculae are better visible from vantage points near the poles during these cycles.

The dependence of the rotational variability on the inclination can be surprisingly well approximated by a simple function of the type
\begin{equation}
    \langle S_{30}\rangle({\rm i}) = {\rm c1}\,+\,{\rm c2}\,{\rm sin(i)}\ ,
\end{equation}
where c1 and c2 are constants. The fit to the rotational variability for the full time series using the above function with ${\rm c1}=0.003$ and ${\rm c2}=0.018$ is shown by the black dotted curve in \fref{fig:rotats}a. The rotational variability normalized to the value at ${\rm i}=90\degree$ for the different activity cycles and the full time series are plotted in \fref{fig:rotats}b. The black dotted curve indicates the fit to the full time series for which ${\rm c1}=0.154$ and ${\rm c2}=0.846$ have been used.

Figure~\ref{fig:cycle}a presents the rotational variability ($\langle S_{30}\rangle$) and the amplitude of the cycle in \sind{} ($\Delta S$) as a function of the solar cycle number, while Figure~\ref{fig:cycle}b presents the rotational variability of \sind{} as a function of its cyclic variability. One can see that the variability on both timescales depends on the strength of the activity cycle. We see a clear correlation between the variability of \sind{} on the rotational and magnetic activity cycle timescales (the linear Pearson correlation coefficient is 0.87).

\begin{figure}[h!]
    \centering
    \includegraphics[trim=0.5cm 0.cm 1.5cm 0.cm,clip,height=12.4cm,width=18cm]{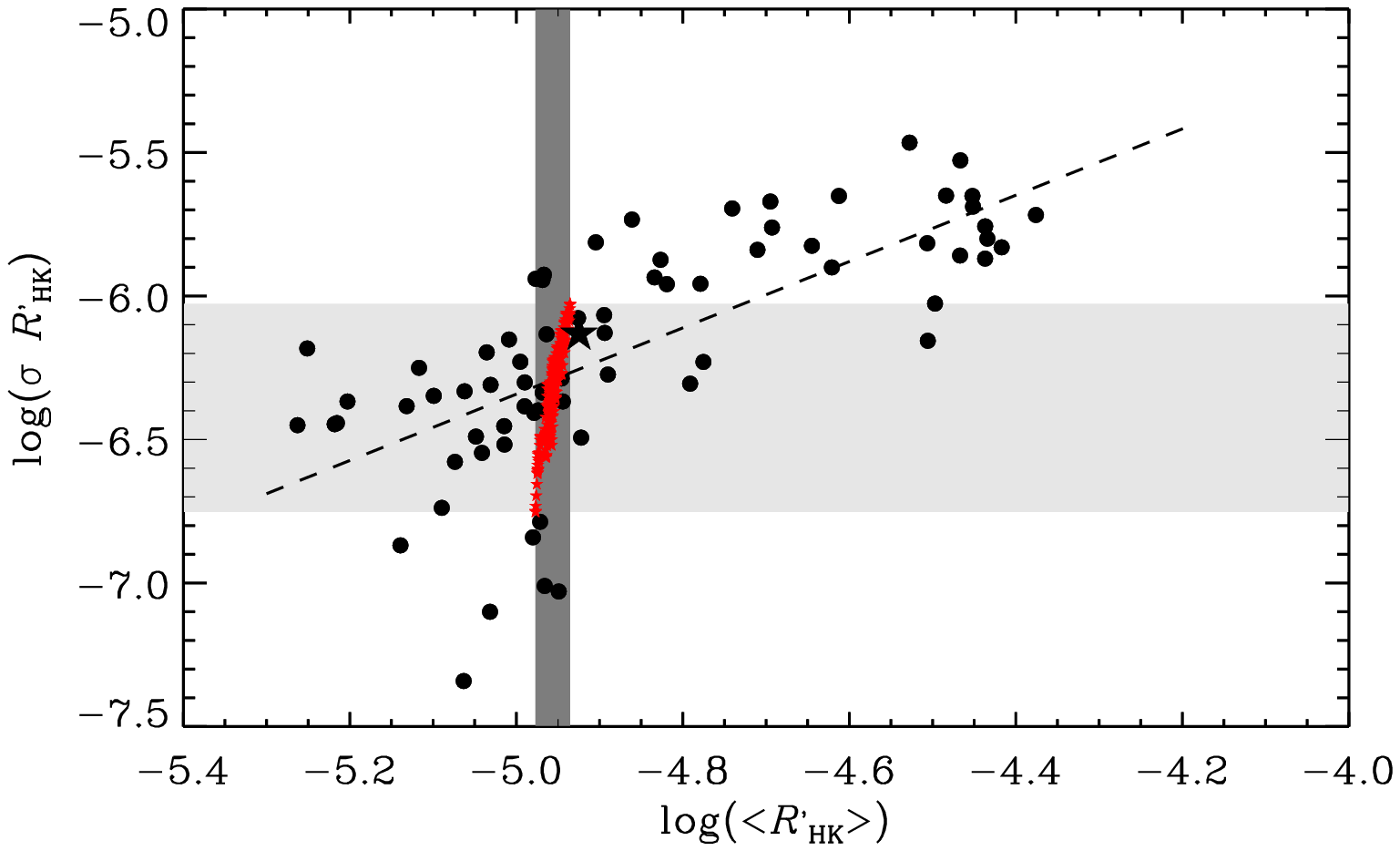}
    \caption{Chromospheric emission variation as a function of mean chromospheric activity. This figure is based on Figure 11a of \citet{2018ApJ...855...75R}. Black star symbol corresponds to the Sun, as marked by \citet{2018ApJ...855...75R} and black circles to other stars in their sample. Overplotted red star symbols are the solar values for all 10 inclinations and for each of the 23 solar cycles. The gray shaded areas highlight the entire range of the variability spanned by all 23 cycles at all inclinations. See \sref{ssec:anomaly} for more details.}
    \label{fig:radick}
\end{figure}

\subsection{Is the Sun anomalous in its chromospheric activity?}
\label{ssec:anomaly}
In this section, we discuss the effect of the inclination on the chromospheric emission variation ($\sigma_S$) and the mean level of the chromospheric activity ($\langle S \rangle$), focusing on the place of the Sun among other Sun-like stars. $\sigma_S$ is defined as the standard deviation of the annual averages of the \sind{} in a given time period and $\langle S \rangle$ as the mean of the annual averages in the \sind{} in that same period. The analysis of Sun-like stars by \citet{1998ApJS..118..239R} indicated a linear dependence (on a log-log scale) of $\sigma_S$ on $\langle S \rangle$. Interestingly, the solar $\sigma_S$ was higher than what the linear dependence suggested \citep[see also][]{2007ApJS..171..260L}. Such a curious behaviour of the Sun was thought to be caused, to a large part, by the amplitude of the \sind{} being calculated for strong solar cycles (namely, cycles 21 and 22). Indeed, when \citet{2018ApJ...855...75R} included the weaker solar cycle 24, the location of the Sun shifted closer to the linear dependence but still remained above it. Recently, \citet{Gomes2020} rekindled the interest in this problem by comparing the solar data from \citet[][obtained during solar cycles 20--23]{2008ApJ...687.1264M} to 1674 FGK stars observed by HARPS. They found that the Sun is located in the high chromospheric emission variability side of the distribution of inactive main-sequence stars. Our model produces the \sind{} time series for all possible inclinations and over a range of solar cycles with different strengths. This enables us to investigate whether the variability of the solar chromospheric emission is, indeed, overly strong in comparison to stars with near-solar magnetic activity.

In \sref{ssec:svaract} we showed that the amplitude of the cycle in \sind{} is affected by the inclination (in particular, the amplitude is found to be strongest for the Sun observed at an inclination of 90\textdegree). Furthermore, our approach allows us to calculate $\sigma_S$ and $\langle S \rangle$ for cycles back to 1755 which in turn allows us to provide the solar cycle amplitude in the \sind{} for cycles of different strength (as given by the sunspot number). This gives us a chance to illustrate how solar $\sigma_S$ and $\langle S \rangle$ change with the inclination and the activity level. The approach that we take to compute these changes is as follows: (i) compute the annual averages of the \sind{} in the period covering solar cycles $1-23$; (ii) for each cycle, compute the mean ($\langle S \rangle$) and the standard deviation ($\sigma_S$) of the annual averages; (iii) repeat the first two steps for all 10 inclinations (we note however that the effect of inclination on $\langle S \rangle$ is very small compared to that on $\sigma_S$); and (iv) transform these values to the $R^\prime_{\rm HK}$ scale following the steps described in \sref{ssec:rec}.

Figure~\ref{fig:radick} shows the chromospheric emission variation as a function of the mean chromospheric activity and is adapted from Figure 11a of \citealt{2018ApJ...855...75R} (in \fref{fig:radick} we have neglected the sample of \citet{2007ApJS..171..260L} originally present in Figure 11a of \citealt{2018ApJ...855...75R} and have added our results obtained following the steps described above). The upper bounds for the gray shaded areas are set by the strongest cycle (solar cycle 19) at the 90\textdegree\ inclination and the lower bounds by the weakest cycle (solar cycle 6) at 0\textdegree\ inclination. It is clear from this figure that the Sun exhibits a wide range of chromospheric emission variations depending on the strength of the activity cycle and the inclination of its rotation axis relative to the observer. For an intermediately strong solar cycle 23, the inclination might change $\sigma_S$ by up to 30\%. Interestingly, depending on the period of observations the Sun could appear below and above stellar regression line (dashed line in \fref{fig:radick}) with the mean solar level being very close to the regression line. This result unambiguously proves that solar chromospheric emission variation is absolutely normal with respect to other stars with near-solar activity.

\section{Conclusions}
\label{sec:concl}
We have developed a physics-based model based on the SATIRE approach to calculate the \sind{} and its variations with solar activity and to quantify the effect of the stellar inclination on the \sind{}. To validate our model, we used the observed surface area coverages of solar faculae from \mbox{SATIRE-S} \citep{2014A&A...570A..85Y} in combination with the non-LTE spectra synthesized from the RH code. Our model reproduced the observed variation of the \sind{} for solar cycles $21-24$ in the composite produced from \citet{2016SoPh..291.2967B} both on the rotational and activity cycle timescales. In particular, we showed that the cycle parameters such as the minimum, maximum, average, and amplitude, calculated with our model are in close agreement with those obtained using the composite of \citet{2016SoPh..291.2967B} and found by \citet{2017ApJ...835...25E}.

We have also studied how the amplitudes of the activity and rotational cycles in the \sind{} depend on stellar inclination. Due to the lack of observations of the Sun out of the ecliptic plane, we made use of the surface distribution of faculae simulated with a SFTM \citep{Nina2,Nina1}. We found that the amplitude of the \sind{} averaged over a moderately strong activity cycle decreases by about 22\%\ when the inclination changes from an equator-on view to a pole-on view. This is due to the decrease in the facular area coverage on the visible disk. In addition, the foreshortening effect dominates over the relatively small increase of the facular contrast in the H and K passbands towards the limb, leading to a reduced \ca{} emission closer to the limb. The amplitude of the rotational variability in the \sind{} decreases much more, by about 81\%\, with decreasing inclination. Both the activity and rotational cycles in the \sind{} depend on the strength of the magnetic activity cycles. We also found that the inclination dependence of the activity cycles in \sind{} can be modeled by a logistic function while the rotational cycles in the \sind{} can be described by a simple linear function that depends on sin(i).

Finally, we investigated the influence of the inclination on the relationship between the mean level and the variation of the chromospheric emission. Such a relationship was found for sun-like stars on the lower main sequence \citep{2018ApJ...855...75R}. To allow a more realistic comparison of the chromospheric activity of the Sun with other sun-like stars which are observed at random inclinations, we calculated \sind{} for 10 inclinations and each of the 23 solar cycles. This helped us to assess the potential range spanned by the chromospheric activity variations, if the Sun were observed at an arbitrary inclination and during an arbitrary activity cycle. We found that depending on the inclination and the period of observations, the activity cycle in solar \sind{} can appear weaker or stronger than in stars with a solar-like level of magnetic activity. We thus conclude that there is nothing unusual about the solar chromospheric emission variations in the context of stars with near-solar magnetic activity.

Further detailed studies are needed to explore the dependence of the \sind{} on the stellar fundamental parameters such as the effective temperature, surface gravity, and metallicity. For example, the effective temperature of a star defines the flux in the pseudo-continuum region and thus directly affects the \sind{}. While this was taken into account when introducing the derivative index $R^\prime_{\rm HK}$ \citep{1984ApJ...279..763N,1984A&A...130..353R}, which makes the comparison of the chromospheric activity of stars of different spectral classes easier, it has not been investigated to what extent a change of the effective temperature and metallicity also affects the radiative transfer in the \ca{} emission and pseudo-continuum. In a future study we plan to extend our model to investigate the effect of stellar fundamental parameters on the intricate relationship between the stellar magnetic activity and the \ca{} emission.

\acknowledgments
We thank an anonymous referee for the constructive comments which helped improve the paper. We thank H.-P.~Doerr for the useful discussions related to FTS Atlas. KS received funding from the European Union's Horizon 2020 research and innovation programme under the Marie Sk{\l}odowska-Curie grant agreement No. 797715. AIS, VW, and NEN have received funding from the European Research Council under the European Union's Horizon 2020 research and innovation program (grant agreement No. 715947). TC and NAK acknowledge support by the German Federal Ministry of Education and Research (Project No. 01LG1909C). SKS has received funding from the European Research Council under the European Union’s Horizon 2020 research and innovation programme (grant agreement No. 695075) and has been supported by the BK21 plus program through the National Research Foundation funded by the Ministry of Education of Korea. We also thank the International Space Science Institute, Bern, for their support of science team 446 and the resulting helpful discussions. This study has made use of SAO/NASA Astrophysics Data System's bibliographic services.

\appendix 
\restartappendixnumbering

\section{Comparison of quiet Sun and plage spectra}
\label{app:qsplage}
\begin{figure*}[ht!]
    \centering
    \includegraphics[scale=0.46]{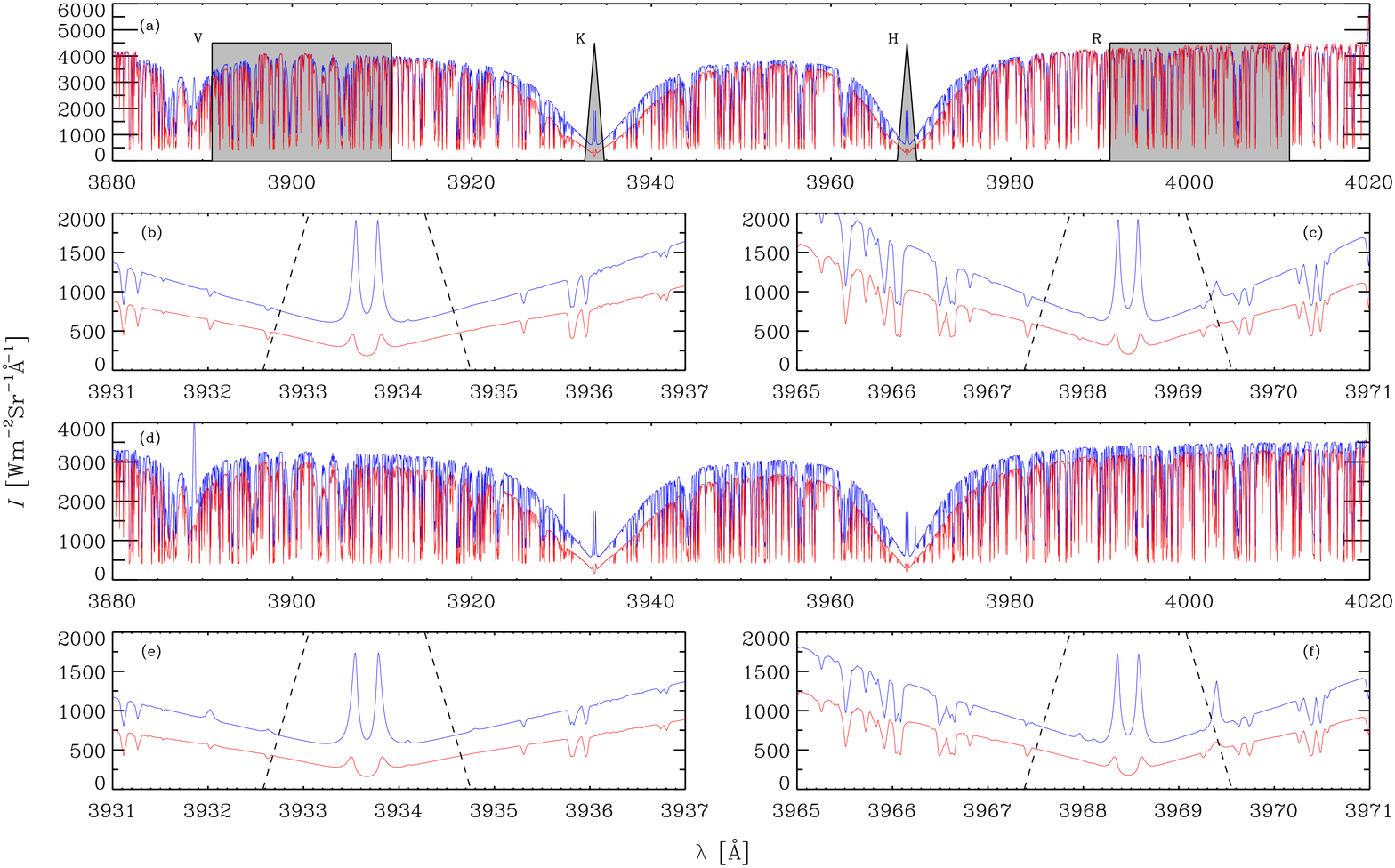}
    \caption{Comparison between the quiet Sun (red) and the plage spectra (blue) synthesized with the RH code. Panels (a), (b), and (c) show the disk-center spectra while the disk-integrated spectra are shown in panels (d), (e), and (f). The gray shaded regions and the dashed lines have the same meaning as in \fref{fig:spectra}.}
    \label{fig:qsplage}
\end{figure*}
Figure~\ref{fig:qsplage} shows a comparison between the plage spectra (blue) and the quiet-Sun spectra (red) calculated for the disk-center (panels a--c) and integrated over the solar disk (panels d--f). The quiet-Sun spectrum is calculated with model FALC99 from \cite{1999ApJ...518..480F} and is identical to that shown in \fref{fig:spectra}.  The plage spectrum is calculated with the FALP99 model from \cite{1999ApJ...518..480F}. The cores of the H and K lines form in the chromosphere and their profiles are substantially different in the two model atmospheres, as can be seen in panels b, c, e, and f \citep[see also][]{1972SoPh...25..357S}. The non-thermal heating due to magnetic fields, empirically taken into account in FALP99, leads to prominent emissions in the line cores (see ${\rm K}_2$ and ${\rm H}_2$ features in plage spectra). The wavelengths in the pseudo-continuum bands form in the photosphere where the two model atmospheres return similar spectra.

The temperature difference between the FALC99 and FALP99 atmospheres increases with height \citep[see also][]{1974SoPh...39...49S}. In the disk-integrated case, the emission forms on average higher in the atmosphere as compared to the disk center. As a result of this, the contrasts are higher for the disk-integrated spectra (see panel d). We note that some background spectral lines in the disk-integrated spectra are seen in emission (e.g. the hydrogen line in panel f at 3969.4\,\AA{}). In our spectral synthesis, these lines are treated in LTE. For smaller $\mu$ values (i.e, closer to the limb), they form in the chromosphere. This leads to an increase in the source function with height thus leading to the spurious emissions. The \sind{} calculations however are not affected by them since such spurious emissions fall outside the passbands of relevance.

\section{Effect of macroturbulence on the quiet Sun and plage spectra}
\label{app:macro}
\begin{figure*}[ht!]
    \centering
    \includegraphics[scale=1.,trim=0.5cm 1.5cm 0.1cm 1.0cm,clip]{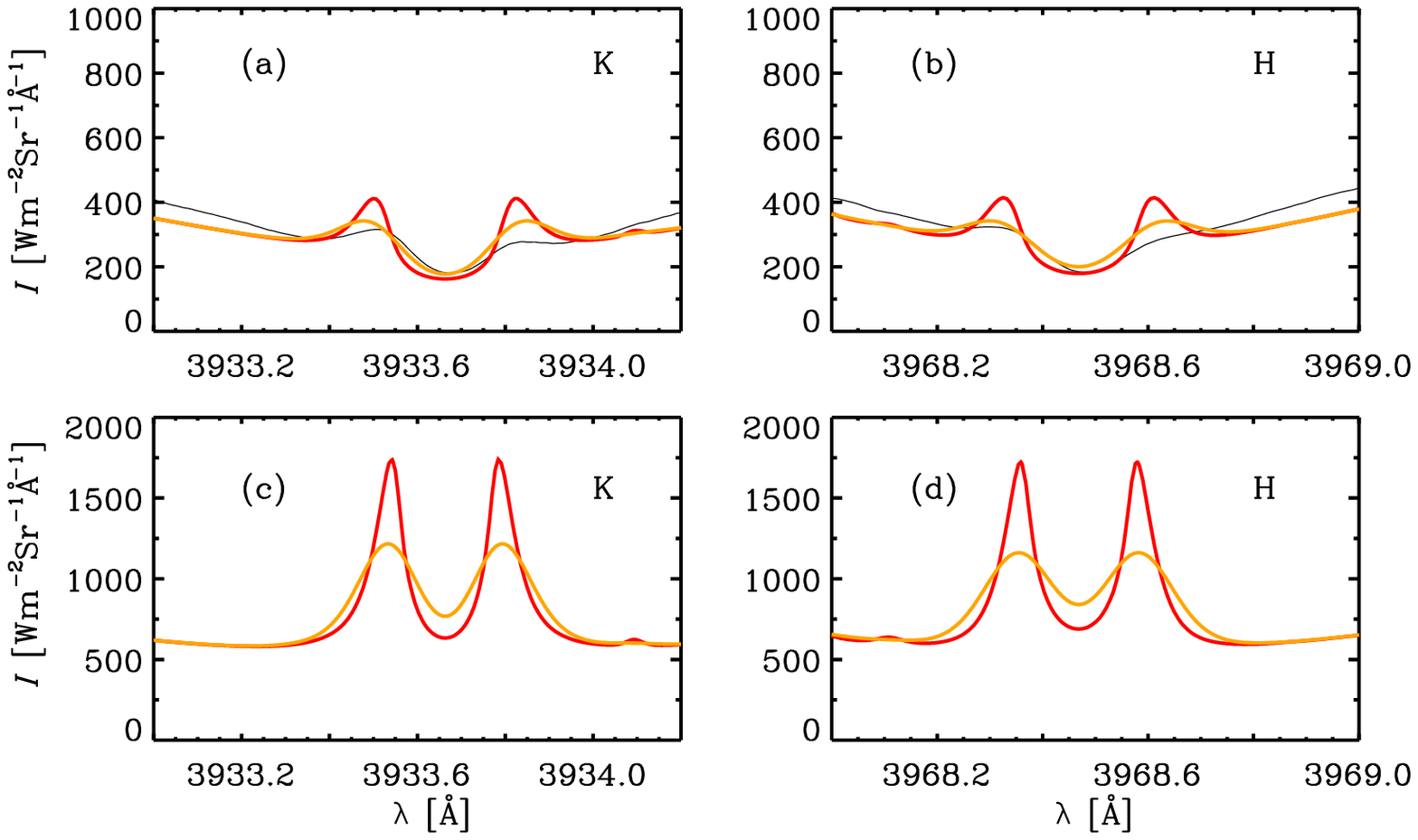}
    \caption{Comparison between the \ca{} profiles before (red) and after (orange) convolving the synthetic spectra using a Gaussian profile with a FWHM of 120\,m\AA{} to account for the broadening due to macroturbulent velocity fields. Panels (a) and (b) show the disk-integrated quiet Sun profiles, with the observed FTS spectra in black, while the disk-integrated plage profiles are shown in panels (c) and (d).}
    \label{fig:macro}
\end{figure*}

Figure~\ref{fig:macro} shows the effect of macroturbulent velocity fields on the \ca{} line profiles. The disk-integrated profiles synthesized using RH are convolved with a Gaussian profile having a FWHM of 120\,m\AA{}. This width corresponds to a macroturbulent velocity of $\sim9$\,kms$^{-1}$. Such velocity fields lead to a broadening of the line core thereby bringing down the enhanced emission in the K$_2$ and H$_2$ features, as illustrated in the figure. The smearing of the \ca{} profiles caused by the macroturbulent velocity fields leads to a better agreement between the synthetic and the observed profiles as shown in panels (a) and (b). However, the effect of macroturbulence on the \sind{} is negligible.

\bibliographystyle{aasjournal}
\bibliography{sindex}

\begin{thebibliography}{}
\expandafter\ifx\csname natexlab\endcsname\relax\def\natexlab#1{#1}\fi
\providecommand{\url}[1]{\href{#1}{#1}}
\providecommand{\dodoi}[1]{doi:~\href{http://doi.org/#1}{\nolinkurl{#1}}}
\providecommand{\doeprint}[1]{\href{http://ascl.net/#1}{\nolinkurl{http://ascl.net/#1}}}
\providecommand{\doarXiv}[1]{\href{https://arxiv.org/abs/#1}{\nolinkurl{https://arxiv.org/abs/#1}}}

\bibitem[{{Anders} \& {Grevesse}(1989)}]{1989GeCoA..53..197A}
{Anders}, E., \& {Grevesse}, N. 1989, \gca, 53, 197,
  \dodoi{10.1016/0016-7037(89)90286-X}

\bibitem[{{Balasubramaniam} \& {Pevtsov}(2011)}]{2011SPIE.8148E..09B}
{Balasubramaniam}, K.~S., \& {Pevtsov}, A. 2011, Society of Photo-Optical
  Instrumentation Engineers (SPIE) Conference Series, Vol. 8148, {Ground-based
  synoptic instrumentation for solar observations} (SPIE), 814809,
  \dodoi{10.1117/12.892824}

\bibitem[{{Baliunas} {et~al.}(1998){Baliunas}, {Donahue}, {Soon}, \&
  {Henry}}]{1998ASPC..154..153B}
{Baliunas}, S.~L., {Donahue}, R.~A., {Soon}, W., \& {Henry}, G.~W. 1998,
  Astronomical Society of the Pacific Conference Series, Vol. 154, {Activity
  Cycles in Lower Main Sequence and POST Main Sequence Stars: The HK Project}
  (ASPC), 153

\bibitem[{{Ball} {et~al.}(2014){Ball}, {Krivova}, {Unruh}, {Haigh}, \&
  {Solanki}}]{balletal2014}
{Ball}, W.~T., {Krivova}, N.~A., {Unruh}, Y.~C., {Haigh}, J.~D., \& {Solanki},
  S.~K. 2014, J. Atmos. Sci., 71, 4086, \dodoi{10.1175/JAS-D-13-0241.1}

\bibitem[{{Basri} {et~al.}(2013){Basri}, {Walkowicz}, \&
  {Reiners}}]{2013ApJ...769...37B}
{Basri}, G., {Walkowicz}, L.~M., \& {Reiners}, A. 2013, \apj, 769, 37,
  \dodoi{10.1088/0004-637X/769/1/37}

\bibitem[{{Bertello} {et~al.}(2016){Bertello}, {Pevtsov}, {Tlatov}, \&
  {Singh}}]{2016SoPh..291.2967B}
{Bertello}, L., {Pevtsov}, A., {Tlatov}, A., \& {Singh}, J. 2016, \solphys,
  291, 2967, \dodoi{10.1007/s11207-016-0927-9}

\bibitem[{{Bj{\o}rgen} {et~al.}(2018){Bj{\o}rgen}, {Sukhorukov}, {Leenaarts},
  {Carlsson}, {de la Cruz Rodr{\'\i}guez}, {Scharmer}, \&
  {Hansteen}}]{2018A&A...611A..62B}
{Bj{\o}rgen}, J.~P., {Sukhorukov}, A.~V., {Leenaarts}, J., {et~al.} 2018, \aap,
  611, A62, \dodoi{10.1051/0004-6361/201731926}

\bibitem[{{Bruls} {et~al.}(1992){Bruls}, {Rutten}, \&
  {Shchukina}}]{1992A&A...265..237B}
{Bruls}, J.~H.~M.~J., {Rutten}, R.~J., \& {Shchukina}, N.~G. 1992, \aap, 265,
  237

\bibitem[{{Cameron} {et~al.}(2010){Cameron}, {Jiang}, {Schmitt}, \&
  {Sch{\"u}ssler}}]{2010ApJ...719..264C}
{Cameron}, R.~H., {Jiang}, J., {Schmitt}, D., \& {Sch{\"u}ssler}, M. 2010,
  \apj, 719, 264, \dodoi{10.1088/0004-637X/719/1/264}

\bibitem[{{Castelli} \& {Kurucz}(1994)}]{1994A&A...281..817C}
{Castelli}, F., \& {Kurucz}, R.~L. 1994, \aap, 281, 817

\bibitem[{Chatzistergos(2017)}]{chatzistergos_analysis_2017}
Chatzistergos, T. 2017, Analysis of historical solar observations and long-term
  changes in solar irradiance, {PhD} thesis (Uni-edition).
\newblock \url{https://ui.adsabs.harvard.edu/abs/2017PhDT.......259C}

\bibitem[{{Chatzistergos} {et~al.}(2017){Chatzistergos}, {Usoskin},
  {Kovaltsov}, {Krivova}, \& {Solanki}}]{2017A&A...602A..69C}
{Chatzistergos}, T., {Usoskin}, I.~G., {Kovaltsov}, G.~A., {Krivova}, N.~A., \&
  {Solanki}, S.~K. 2017, \aap, 602, A69, \dodoi{10.1051/0004-6361/201630045}

\bibitem[{{Cui} {et~al.}(2012){Cui}, {Zhao}, {Chu}, {Li}, {Li}, {Zhang}, {Su},
  {Yao}, {Wang}, {Xing}, {Li}, {Zhu}, {Wang}, {Gu}, {Luo}, {Xu}, {Zhang},
  {Liu}, {Zhang}, {Yang}, {Cao}, {Chen}, {Chen}, {Chen}, {Chen}, {Chu}, {Feng},
  {Gong}, {Hou}, {Hu}, {Hu}, {Hu}, {Jia}, {Jiang}, {Jiang}, {Jiang}, {Jin},
  {Li}, {Li}, {Li}, {Liu}, {Liu}, {Lu}, {Mao}, {Men}, {Qi}, {Qi}, {Shi},
  {Tang}, {Tao}, {Wang}, {Wang}, {Wang}, {Wang}, {Wang}, {Wang}, {Wang},
  {Wang}, {Wang}, {Wang}, {Wang}, {Wang}, {Xu}, {Xu}, {Yang}, {Yu}, {Yuan},
  {Yuan}, {Zhai}, {Zhang}, {Zhang}, {Zhang}, {Zhao}, {Zhou}, {Zhou}, {Zhu}, \&
  {Zou}}]{2012RAA....12.1197C}
{Cui}, X.-Q., {Zhao}, Y.-H., {Chu}, Y.-Q., {et~al.} 2012, Research in Astronomy
  and Astrophysics, 12, 1197, \dodoi{10.1088/1674-4527/12/9/003}

\bibitem[{{Danilovic} {et~al.}(2016){Danilovic}, {Solanki}, {Livingston},
  {Krivova}, \& {Vince}}]{Sanja2016}
{Danilovic}, S., {Solanki}, S.~K., {Livingston}, W., {Krivova}, N., \& {Vince},
  I. 2016, \aap, 587, A33, \dodoi{10.1051/0004-6361/201527039}

\bibitem[{{Dasi-Espuig} {et~al.}(2016){Dasi-Espuig}, {Jiang}, {Krivova},
  {Solanki}, {Unruh}, \& {Yeo}}]{Dasi2016}
{Dasi-Espuig}, M., {Jiang}, J., {Krivova}, N.~A., {et~al.} 2016, \aap, 590,
  A63, \dodoi{10.1051/0004-6361/201527993}

\bibitem[{{De Cat} {et~al.}(2015){De Cat}, {Fu}, {Ren}, {Yang}, {Shi}, {Luo},
  {Yang}, {Wang}, {Zhang}, {Shi}, {Zhang}, {Dong}, {Catanzaro}, {Corbally},
  {Frasca}, {Gray}, {Molenda-{\.Z}akowicz}, {Uytterhoeven}, {Briquet},
  {Bruntt}, {Frandsen}, {Kiss}, {Kurtz}, {Marconi}, {Niemczura}, {{\O}stensen},
  {Ripepi}, {Smalley}, {Southworth}, {Szab{\'o}}, {Telting}, {Karoff}, {Silva
  Aguirre}, {Wu}, {Hou}, {Jin}, \& {Zhou}}]{2015ApJS..220...19D}
{De Cat}, P., {Fu}, J.~N., {Ren}, A.~B., {et~al.} 2015, \apjs, 220, 19,
  \dodoi{10.1088/0067-0049/220/1/19}

\bibitem[{{Doerr} {et~al.}(2016){Doerr}, {Vitas}, \&
  {Fabbian}}]{2016A&A...590A.118D}
{Doerr}, H.~P., {Vitas}, N., \& {Fabbian}, D. 2016, \aap, 590, A118,
  \dodoi{10.1051/0004-6361/201628570}

\bibitem[{{Du}(2011)}]{2011SoPh..273..231D}
{Du}, Z. 2011, \solphys, 273, 231, \dodoi{10.1007/s11207-011-9849-8}

\bibitem[{{Duncan} {et~al.}(1991){Duncan}, {Vaughan}, {Wilson}, {Preston},
  {Frazer}, {Lanning}, {Misch}, {Mueller}, {Soyumer}, {Woodard}, {Baliunas},
  {Noyes}, {Hartmann}, {Porter}, {Zwaan}, {Middelkoop}, {Rutten}, \&
  {Mihalas}}]{1991ApJS...76..383D}
{Duncan}, D.~K., {Vaughan}, A.~H., {Wilson}, O.~C., {et~al.} 1991, \apjs, 76,
  383, \dodoi{10.1086/191572}

\bibitem[{{Egeland} {et~al.}(2017{\natexlab{a}}){Egeland}, {Soon}, {Baliunas},
  {Hall}, {Pevtsov}, \& {Bertello}}]{2017ApJ...835...25E}
{Egeland}, R., {Soon}, W., {Baliunas}, S., {et~al.} 2017{\natexlab{a}}, \apj,
  835, 25, \dodoi{10.3847/1538-4357/835/1/25}

\bibitem[{{Egeland} {et~al.}(2017{\natexlab{b}}){Egeland}, {Soon}, {Baliunas},
  {Hall}, {Pevtsov}, \& {Bertello}}]{2017yCat..18350025E}
---. 2017{\natexlab{b}}, VizieR Online Data Catalog, J/ApJ/835/25

\bibitem[{{Engvold} {et~al.}(2019){Engvold}, {Vial}, \&
  {Skumanich}}]{best_book_ever}
{Engvold}, O., {Vial}, J.-C., \& {Skumanich}, A. 2019, {The Sun as a Guide to
  Stellar Physics} (Elsevier), \dodoi{10.1016/C2017-0-01365-4}

\bibitem[{{Fligge} {et~al.}(2000){Fligge}, {Solanki}, \&
  {Unruh}}]{2000A&A...353..380F}
{Fligge}, M., {Solanki}, S.~K., \& {Unruh}, Y.~C. 2000, \aap, 353, 380

\bibitem[{{Fontenla} {et~al.}(1999){Fontenla}, {White}, {Fox}, {Avrett}, \&
  {Kurucz}}]{1999ApJ...518..480F}
{Fontenla}, J., {White}, O.~R., {Fox}, P.~A., {Avrett}, E.~H., \& {Kurucz},
  R.~L. 1999, \apj, 518, 480, \dodoi{10.1086/307258}

\bibitem[{{Gandorfer} {et~al.}(2011){Gandorfer}, {Grauf}, {Barthol},
  {Riethm{\"u}ller}, {Solanki}, {Chares}, {Deutsch}, {Ebert}, {Feller},
  {Germerott}, {Heerlein}, {Heinrichs}, {Hirche}, {Hirzberger}, {Kolleck},
  {Meller}, {M{\"u}ller}, {Sch{\"a}fer}, {Tomasch}, {Kn{\"o}lker},
  {Mart{\'\i}nez Pillet}, {Bonet}, {Schmidt}, {Berkefeld}, {Feger}, {Heidecke},
  {Soltau}, {Tischenberg}, {Fischer}, {Title}, {Anwand}, \&
  {Schmidt}}]{2011SoPh..268...35G}
{Gandorfer}, A., {Grauf}, B., {Barthol}, P., {et~al.} 2011, \solphys, 268, 35,
  \dodoi{10.1007/s11207-010-9636-y}

\bibitem[{{Gomes da Silva} {et~al.}(2020){Gomes da Silva}, {Santos},
  {Adibekyan}, {Sousa}, {Campante}, {Figueira}, {Bossini}, {Delgado-Mena},
  {Monteiro}, {de Lavern}, {Recio-Blanco}, \& {Lovis}}]{Gomes2020}
{Gomes da Silva}, J., {Santos}, N.~C., {Adibekyan}, V., {et~al.} 2020, arXiv
  e-prints, arXiv:2012.10199.
\newblock \doarXiv{2012.10199}

\bibitem[{{Hall} {et~al.}(2007){Hall}, {Lockwood}, \&
  {Skiff}}]{2007AJ....133..862H}
{Hall}, J.~C., {Lockwood}, G.~W., \& {Skiff}, B.~A. 2007, \aj, 133, 862,
  \dodoi{10.1086/510356}

\bibitem[{{Hathaway}(2015)}]{2015LRSP...12....4H}
{Hathaway}, D.~H. 2015, Living Reviews in Solar Physics, 12, 4,
  \dodoi{10.1007/lrsp-2015-4}

\bibitem[{{Hathaway} {et~al.}(1994){Hathaway}, {Wilson}, \&
  {Reichmann}}]{1994SoPh..151..177H}
{Hathaway}, D.~H., {Wilson}, R.~M., \& {Reichmann}, E.~J. 1994, \solphys, 151,
  177, \dodoi{10.1007/BF00654090}

\bibitem[{{Jafarzadeh} {et~al.}(2017){Jafarzadeh}, {Rutten}, {Solanki},
  {Wiegelmann}, {Riethm{\"u}ller}, {van Noort}, {Szydlarski}, {Blanco
  Rodr{\'\i}guez}, {Barthol}, {del Toro Iniesta}, {Gandorfer}, {Gizon},
  {Hirzberger}, {Kn{\"o}lker}, {Mart{\'\i}nez Pillet}, {Orozco Su{\'a}rez}, \&
  {Schmidt}}]{2017ApJS..229...11J}
{Jafarzadeh}, S., {Rutten}, R.~J., {Solanki}, S.~K., {et~al.} 2017, \apjs, 229,
  11, \dodoi{10.3847/1538-4365/229/1/11}

\bibitem[{{Jiang} {et~al.}(2011){Jiang}, {Cameron}, {Schmitt}, \&
  {Sch{\"u}ssler}}]{2011A&A...528A..82J}
{Jiang}, J., {Cameron}, R.~H., {Schmitt}, D., \& {Sch{\"u}ssler}, M. 2011,
  \aap, 528, A82, \dodoi{10.1051/0004-6361/201016167}

\bibitem[{{Keil} \& {Worden}(1984)}]{1984ApJ...276..766K}
{Keil}, S.~L., \& {Worden}, S.~P. 1984, \apj, 276, 766, \dodoi{10.1086/161663}

\bibitem[{{Knaack} {et~al.}(2001){Knaack}, {Fligge}, {Solanki}, \&
  {Unruh}}]{2001A&A...376.1080K}
{Knaack}, R., {Fligge}, M., {Solanki}, S.~K., \& {Unruh}, Y.~C. 2001, \aap,
  376, 1080, \dodoi{10.1051/0004-6361:20011007}

\bibitem[{{Krivova} {et~al.}(2003){Krivova}, {Solanki}, {Fligge}, \&
  {Unruh}}]{2003A&A...399L...1K}
{Krivova}, N.~A., {Solanki}, S.~K., {Fligge}, M., \& {Unruh}, Y.~C. 2003, \aap,
  399, L1, \dodoi{10.1051/0004-6361:20030029}

\bibitem[{{Krivova} {et~al.}(2011){Krivova}, {Solanki}, \&
  {Unruh}}]{krivova2011}
{Krivova}, N.~A., {Solanki}, S.~K., \& {Unruh}, Y.~C. 2011, Journal of
  Atmospheric and Solar-Terrestrial Physics, 73, 223,
  \dodoi{10.1016/j.jastp.2009.11.013}

\bibitem[{{Kurucz}(1992)}]{1992RMxAA..23...45K}
{Kurucz}, R.~L. 1992, \rmxaa, 23, 45

\bibitem[{{Kurucz}(2009)}]{2009AIPC.1171...43K}
{Kurucz}, R.~L. 2009, in American Institute of Physics Conference Series, Vol.
  1171, American Institute of Physics Conference Series, ed. I.~{Hubeny}, J.~M.
  {Stone}, K.~{MacGregor}, \& K.~{Werner}, 43--51, \dodoi{10.1063/1.3250087}

\bibitem[{{Linsky} \& {Avrett}(1970)}]{1970PASP...82..169L}
{Linsky}, J.~L., \& {Avrett}, E.~H. 1970, \pasp, 82, 169,
  \dodoi{10.1086/128904}

\bibitem[{{Lockwood} {et~al.}(2007){Lockwood}, {Skiff}, {Henry}, {Henry},
  {Radick}, {Baliunas}, {Donahue}, \& {Soon}}]{2007ApJS..171..260L}
{Lockwood}, G.~W., {Skiff}, B.~A., {Henry}, G.~W., {et~al.} 2007, \apjs, 171,
  260, \dodoi{10.1086/516752}

\bibitem[{{Loukitcheva} {et~al.}(2017){Loukitcheva}, {Iwai}, {Solanki},
  {White}, \& {Shimojo}}]{2017ApJ...850...35L}
{Loukitcheva}, M.~A., {Iwai}, K., {Solanki}, S.~K., {White}, S.~M., \&
  {Shimojo}, M. 2017, \apj, 850, 35, \dodoi{10.3847/1538-4357/aa91cc}

\bibitem[{{Mamajek} \& {Hillenbrand}(2008)}]{2008ApJ...687.1264M}
{Mamajek}, E.~E., \& {Hillenbrand}, L.~A. 2008, \apj, 687, 1264,
  \dodoi{10.1086/591785}

\bibitem[{{Mu{\~n}oz-Jaramillo} \& {Vaquero}(2019)}]{2019NatAs...3..205M}
{Mu{\~n}oz-Jaramillo}, A., \& {Vaquero}, J.~M. 2019, Nature Astronomy, 3, 205,
  \dodoi{10.1038/s41550-018-0638-2}

\bibitem[{{Neckel}(1999)}]{1999SoPh..184..421N}
{Neckel}, H. 1999, \solphys, 184, 421, \dodoi{10.1023/A:1017165208013}

\bibitem[{{Neckel} \& {Labs}(1984)}]{1984SoPh...90..205N}
{Neckel}, H., \& {Labs}, D. 1984, \solphys, 90, 205, \dodoi{10.1007/BF00173953}

\bibitem[{{N{\`e}mec} {et~al.}(2020{\natexlab{a}}){N{\`e}mec},
  {I{\textcommabelow s}{\i}k}, {Shapiro}, {Solanki}, {Krivova}, \&
  {Unruh}}]{Nina2}
{N{\`e}mec}, N.~E., {I{\textcommabelow s}{\i}k}, E., {Shapiro}, A.~I., {et~al.}
  2020{\natexlab{a}}, \aap, 638, A56, \dodoi{10.1051/0004-6361/202038054}

\bibitem[{{N{\`e}mec} {et~al.}(2020{\natexlab{b}}){N{\`e}mec}, {Shapiro},
  {Krivova}, {Solanki}, {Tagirov}, {Cameron}, \& {Dreizler}}]{Nina1}
{N{\`e}mec}, N.~E., {Shapiro}, A.~I., {Krivova}, N.~A., {et~al.}
  2020{\natexlab{b}}, \aap, 636, A43, \dodoi{10.1051/0004-6361/202037588}

\bibitem[{{Noyes} {et~al.}(1984){Noyes}, {Hartmann}, {Baliunas}, {Duncan}, \&
  {Vaughan}}]{1984ApJ...279..763N}
{Noyes}, R.~W., {Hartmann}, L.~W., {Baliunas}, S.~L., {Duncan}, D.~K., \&
  {Vaughan}, A.~H. 1984, \apj, 279, 763, \dodoi{10.1086/161945}

\bibitem[{{Pneuman} {et~al.}(1986){Pneuman}, {Solanki}, \&
  {Stenflo}}]{1986A&A...154..231P}
{Pneuman}, G.~W., {Solanki}, S.~K., \& {Stenflo}, J.~O. 1986, \aap, 154, 231

\bibitem[{{Radick} {et~al.}(2018){Radick}, {Lockwood}, {Henry}, {Hall}, \&
  {Pevtsov}}]{2018ApJ...855...75R}
{Radick}, R.~R., {Lockwood}, G.~W., {Henry}, G.~W., {Hall}, J.~C., \&
  {Pevtsov}, A.~A. 2018, \apj, 855, 75, \dodoi{10.3847/1538-4357/aaaae3}

\bibitem[{{Radick} {et~al.}(1998){Radick}, {Lockwood}, {Skiff}, \&
  {Baliunas}}]{1998ApJS..118..239R}
{Radick}, R.~R., {Lockwood}, G.~W., {Skiff}, B.~A., \& {Baliunas}, S.~L. 1998,
  \apjs, 118, 239, \dodoi{10.1086/313135}

\bibitem[{{Rutten}(1984)}]{1984A&A...130..353R}
{Rutten}, R.~G.~M. 1984, \aap, 130, 353

\bibitem[{{Rutten}(2019)}]{2019SoPh..294..165R}
{Rutten}, R.~J. 2019, \solphys, 294, 165, \dodoi{10.1007/s11207-019-1535-2}

\bibitem[{{Rybicki} \& {Hummer}(1991)}]{1991A&A...245..171R}
{Rybicki}, G.~B., \& {Hummer}, D.~G. 1991, \aap, 245, 171

\bibitem[{{Rybicki} \& {Hummer}(1992)}]{1992A&A...262..209R}
---. 1992, \aap, 262, 209

\bibitem[{{Schatten}(1993)}]{1993JGR....9818907S}
{Schatten}, K.~H. 1993, \jgr, 98, 18907, \dodoi{10.1029/93JA01941}

\bibitem[{{Shapiro} {et~al.}(2014){Shapiro}, {Solanki}, {Krivova}, {Schmutz},
  {Ball}, {Knaack}, {Rozanov}, \& {Unruh}}]{2014A&A...569A..38S}
{Shapiro}, A.~I., {Solanki}, S.~K., {Krivova}, N.~A., {et~al.} 2014, \aap, 569,
  A38, \dodoi{10.1051/0004-6361/201323086}

\bibitem[{{Shine} \& {Linsky}(1972)}]{1972SoPh...25..357S}
{Shine}, R.~A., \& {Linsky}, J.~L. 1972, \solphys, 25, 357,
  \dodoi{10.1007/BF00192335}

\bibitem[{{Shine} \& {Linsky}(1974)}]{1974SoPh...39...49S}
---. 1974, \solphys, 39, 49, \dodoi{10.1007/BF00154970}

\bibitem[{{Solanki} {et~al.}(1999){Solanki}, {Finsterle}, {R{\"u}edi}, \&
  {Livingston}}]{1999A&A...347L..27S}
{Solanki}, S.~K., {Finsterle}, W., {R{\"u}edi}, I., \& {Livingston}, W. 1999,
  \aap, 347, L27

\bibitem[{{Solanki} {et~al.}(2006){Solanki}, {Inhester}, \&
  {Sch{\"u}ssler}}]{Sami_B}
{Solanki}, S.~K., {Inhester}, B., \& {Sch{\"u}ssler}, M. 2006, Rep. Progr.
  Phys., 69, 563, \dodoi{10.1088/0034-4885/69/3/R02}

\bibitem[{{Solanki} \& {Steiner}(1990)}]{1990A&A...234..519S}
{Solanki}, S.~K., \& {Steiner}, O. 1990, \aap, 234, 519

\bibitem[{{Solanki} {et~al.}(1991){Solanki}, {Steiner}, \&
  {Uitenbroeck}}]{1991A&A...250..220S}
{Solanki}, S.~K., {Steiner}, O., \& {Uitenbroeck}, H. 1991, \aap, 250, 220

\bibitem[{{Solanki} {et~al.}(2008){Solanki}, {Wenzler}, \&
  {Schmitt}}]{2008A&A...483..623S}
{Solanki}, S.~K., {Wenzler}, T., \& {Schmitt}, D. 2008, \aap, 483, 623,
  \dodoi{10.1051/0004-6361:20054282}

\bibitem[{{Solanki} {et~al.}(2010){Solanki}, {Barthol}, {Danilovic}, {Feller},
  {Gandorfer}, {Hirzberger}, {Riethm{\"u}ller}, {Sch{\"u}ssler}, {Bonet},
  {Mart{\'\i}nez Pillet}, {del Toro Iniesta}, {Domingo}, {Palacios},
  {Kn{\"o}lker}, {Bello Gonz{\'a}lez}, {Berkefeld}, {Franz}, {Schmidt}, \&
  {Title}}]{2010ApJ...723L.127S}
{Solanki}, S.~K., {Barthol}, P., {Danilovic}, S., {et~al.} 2010, \apjl, 723,
  L127, \dodoi{10.1088/2041-8205/723/2/L127}

\bibitem[{{Spruit}(1976)}]{1976SoPh...50..269S}
{Spruit}, H.~C. 1976, \solphys, 50, 269, \dodoi{10.1007/BF00155292}

\bibitem[{{Tagirov} {et~al.}(2019){Tagirov}, {Shapiro}, {Krivova}, {Unruh},
  {Yeo}, \& {Solanki}}]{Rinat_2019}
{Tagirov}, R.~V., {Shapiro}, A.~I., {Krivova}, N.~A., {et~al.} 2019, \aap, 631,
  A178, \dodoi{10.1051/0004-6361/201935121}

\bibitem[{{Topka} {et~al.}(1992){Topka}, {Tarbell}, \&
  {Title}}]{1992ApJ...396..351T}
{Topka}, K.~P., {Tarbell}, T.~D., \& {Title}, A.~M. 1992, \apj, 396, 351,
  \dodoi{10.1086/171721}

\bibitem[{{Uitenbroek}(1990)}]{1990ASPC....9..103U}
{Uitenbroek}, H. 1990, Astronomical Society of the Pacific Conference Series,
  Vol.~9, {The Solar CAII Lines} (ASPC), 103

\bibitem[{{Uitenbroek}(2001)}]{2001ApJ...557..389U}
---. 2001, \apj, 557, 389, \dodoi{10.1086/321659}

\bibitem[{{Unruh} {et~al.}(1999){Unruh}, {Solanki}, \&
  {Fligge}}]{1999A&A...345..635U}
{Unruh}, Y.~C., {Solanki}, S.~K., \& {Fligge}, M. 1999, \aap, 345, 635

\bibitem[{{Vaughan} {et~al.}(1978){Vaughan}, {Preston}, \&
  {Wilson}}]{1978PASP...90..267V}
{Vaughan}, A.~H., {Preston}, G.~W., \& {Wilson}, O.~C. 1978, \pasp, 90, 267,
  \dodoi{10.1086/130324}

\bibitem[{{Vernazza} {et~al.}(1981){Vernazza}, {Avrett}, \&
  {Loeser}}]{1981ApJS...45..635V}
{Vernazza}, J.~E., {Avrett}, E.~H., \& {Loeser}, R. 1981, \apjs, 45, 635,
  \dodoi{10.1086/190731}

\bibitem[{{White} \& {Livingston}(1978)}]{1978ApJ...226..679W}
{White}, O.~R., \& {Livingston}, W. 1978, \apj, 226, 679,
  \dodoi{10.1086/156650}

\bibitem[{{Wilson}(1978)}]{1978ApJ...226..379W}
{Wilson}, O.~C. 1978, \apj, 226, 379, \dodoi{10.1086/156618}

\bibitem[{{Witzke} {et~al.}(2018){Witzke}, {Shapiro}, {Solanki}, {Krivova}, \&
  {Schmutz}}]{Veronika_2018}
{Witzke}, V., {Shapiro}, A.~I., {Solanki}, S.~K., {Krivova}, N.~A., \&
  {Schmutz}, W. 2018, \aap, 619, A146, \dodoi{10.1051/0004-6361/201833936}

\bibitem[{{Yeo} {et~al.}(2014){Yeo}, {Krivova}, {Solanki}, \&
  {Glassmeier}}]{2014A&A...570A..85Y}
{Yeo}, K.~L., {Krivova}, N.~A., {Solanki}, S.~K., \& {Glassmeier}, K.~H. 2014,
  \aap, 570, A85, \dodoi{10.1051/0004-6361/201423628}

\bibitem[{{Zhang} {et~al.}(2020){Zhang}, {Bi}, {Li}, {Jiang}, {Li}, {He}, {Yu},
  {Khanna}, {Ge}, {Liu}, {Tian}, {Wu}, \& {Zhang}}]{2020ApJS..247....9Z}
{Zhang}, J., {Bi}, S., {Li}, Y., {et~al.} 2020, \apjs, 247, 9,
  \dodoi{10.3847/1538-4365/ab6165}

\bibitem[{{Zhao} {et~al.}(2012){Zhao}, {Zhao}, {Chu}, {Jing}, \&
  {Deng}}]{2012RAA....12..723Z}
{Zhao}, G., {Zhao}, Y.-H., {Chu}, Y.-Q., {Jing}, Y.-P., \& {Deng}, L.-C. 2012,
  Research in Astronomy and Astrophysics, 12, 723,
  \dodoi{10.1088/1674-4527/12/7/002}

\end{thebibliography}

\end{document}